# A Mathematical Model for Astrocytes Mediated LTP at Single Hippocampal Synapses

Shivendra Tewari

shivendra@isibang.ac.in

*Systems Science and Informatics Unit*
*Indian Statistical Institute*
*8th Mile, Mysore Road*
*Bangalore 560059, India*

Kaushik Majumdar

kmajumdar@isibang.ac.in

*Systems Science and Informatics Unit*
*Indian Statistical Institute*
*8th Mile, Mysore Road*
*Bangalore 560059, India*

Many contemporary studies have shown that astrocytes play a significant role in modulating both short and long form of synaptic plasticity. There are very few experimental models which elucidate the role of astrocyte over Long-term Potentiation (LTP). Recently, Perea & Araque (2007) demonstrated a role of astrocytes in induction of LTP at single hippocampal synapses. They suggested a purely pre-synaptic basis for induction of this N-methyl-D-Aspartate (NMDA) Receptor-independent LTP. Also, the mechanisms underlying this pre-synaptic induction were not investigated. Here, in this article, we propose a mathematical model for astrocyte modulated LTP which successfully imitates the experimental findings of Perea & Araque (2007). Our study suggests the role of retrograde messengers, possibly Nitric Oxide (NO), for this pre-synaptically modulated LTP.

## 1. Introduction

Long-term Potentiation (LTP), a long form of synaptic plasticity, has long been considered an eminent synaptic model for investigating the molecular basis of memory (Bliss & Collingridge, 1993; Collingridge et al., 2004). However, there is still debate regarding its functional relevance – for example, is it a model of memory formation or is it the exact mechanism used by brain to store information (Bliss et al., 2004)? LTP can be induced using different protocols of stimulation, like, theta-burst stimulation (TBS) protocol or primed-burst stimulation (PBS) protocol (Bliss & Collingridge, 1993). The significance of these protocols is that similar synchronized firing patterns occur inside hippocampus during learning. LTP can be classified depending upon the need for protein transcription and translation for its induction. LTP which does not require protein transcription and translation is termed as early-phase LTP (e-LTP) and lasts for 2 – 4 hrs, whereas, LTP which requires protein transcription and translation is termed as late-phase LTP (l-LTP) and lasts for hours, days or even weeks (Malenka & Bear, 2004; Aslam et al., 2009). There is vast amount of literature available over LTP for different experimental models (Bliss & Lomo, 1973; Bliss & Collingridge, 1993; Malenka & Bear, 2004; Anwyl, 2009; Aslam et al., 2009).

With the advent of new imaging techniques, it became possible to discover the role of a subtype of glial cells widely distributed in Central Nervous System (CNS): astrocyte (Ben Achour et al., 2010). Araque et al. (1999) coined the term "Tripartite Synapse (TpS)" which considers a peri-synaptic astrocyte, in addition to our classical synapse, of a pre-synaptic neuron and post-synaptic neuron. However, due to spatio-temporal limitations of current imaging techniques it is not easy to study LTP at a specific TpS. Consequently, such experimental observations are few, of relatively recent origin (Perea & Araque, 2007; Henneberger et al., 2010) and have remained controversial at least to some extent (Agulhon et al., 2010). In such situations mathematical modeling can be a viable alternative to have hitherto unobserved insights.

Perea & Araque (2007) investigated the consequence of astrocyte calcium ($Ca^{2+}$) elevations at single hippocampal synapses, in particular Schaffer collateral – CA1 pyramidal cell (SC – CA1) synapse. They loaded astrocytes with $Ca^{2+}$ - cage o-nitrophenyl-EGTA (NP-EGTA) stimulated selectively using ultraviolet (UV) – flash photolysis, while simultaneously stimulating SC using minimal stimulation protocol which activates single or few synapses (Perea & Araque, 2007). They reported a transient enhancement of probability of neurotransmitter release (Pr), when an astrocyte was stimulated (but not without), accompanied by a transient increase in synaptic efficacy (given as a measure of mean Excitatory Post-Synaptic Current (EPSC) amplitudes, including successes and failures). Further, when they paired astrocyte stimulation with mild depolarization (to -30 mV) of post-synaptic neuron, found persistent increase in Pr, with synaptic efficacy also following likewise, signifying astrocyte modulated LTP.

The LTP observed in their experiments can be classified as e-LTP which has a pre-synaptic locus of induction, because they observed persistent increase in synaptic efficacy along with no-change in synaptic potency (given as a measure of mean EPSC amplitudes of successes). Also, using antagonists, D-2-amino-5-phosphonopentanoic acid (D-AP5), of NMDARs and antagonists, MPEP + LY367385, of group I metabotropic glutamate receptors (mGluRs) they confirmed that the LTP observed was NMDAR-independent with pre-synaptic locus of induction which is more challenging to our vision of classical LTP in CA1 region (Perea & Araque, 2007). Here, in this article we present, a very comprehensive computational model of TpS which provides a mathematical framework for astrocyte modulated NMDAR-independent LTP observed during the experiments of Perea & Araque (2007). There exists another model (Nadkarni et al, 2008) demonstrating the role of astrocytes on synaptic potentiation, but at that time a lot of important details were either undetermined (e.g., readily releasable pool size in astrocytes which has been determined recently by Malarkey & Parpura (2011)) or not modeled by them (e.g. post-synaptic neuron dynamics). A brief comparison between Nadkarni et al (2008) and the proposed model is listed in Table 1. Nadkarni et al (2008) made a commendable effort to reproduce the experimental findings of Kang et al (1998) by coupling astrocyte $Ca^{2+}$ activity with CA3 pyramidal neuron bouton through a heuristic coupling parameter. However, they did not take care of some important biological detail while modeling the TpS (see Appendix A where some of the aspects have been highlighted and compared with our proposed model).

Table 1: A Comparison between Nadkarni et al (2008) and our proposed model

| Signaling Processes Modeled | Nadkarni et al, 2008 | Proposed Model |
|---|---|---|
| Bouton $Ca^{2+}$ | Yes | Yes |
| Bouton $IP_3$ | No | Yes |
| Synaptic Vesicle / Glutamate | Yes / No | Yes / Yes |
| Astrocytic $Ca^{2+}$ | Yes | Yes |
| Astrocytic $IP_3$ | Yes | Yes |
| Extra-synaptic Vesicle / Glutamate | No / No | Yes / Yes |
| Post-Synaptic Current / Potential | Yes / No | Yes / Yes |
| Post-Synaptic Calcium / Buffer | No / No | Yes / Yes |
| Retrograde Signaling | No | Yes |

## 2. The Model

As mentioned in the previous section, the LTP we were trying to model is an NMDAR-independent LTP with purely pre-synaptic locus of induction. The exact mechanism underlying this phenomenon was not investigated by Perea & Araque (2007). However, as demonstrated by a persistent increase in Pr a plausible role of a retrograde messenger, possibly NO is suspected. NO is a retrograde messenger involved in different forms of synaptic plasticity (Garthwaite & Boulton, 1995; Bon & Garthwaite, 2003; Hopper & Garthwaite, 2006; Phillips et al, 2008; Taqatqeh et al, 2009) that may be involved in the astrocyte-induced synaptic potentiation (Alfonso Araque, personal communication). Here, in this section we present the mathematical model associated with the addressed biological problem, whose computational implementation will be presented in the section that follows. Since, Perea & Araque (2007) performed their experiments over immature wistar rats, attempt has been made to develop the model and collect data while keeping in mind immature cells. The mathematical formulations have been described in the subsequent subsections. The major steps involved are presented as a flowchart in Figure 1.

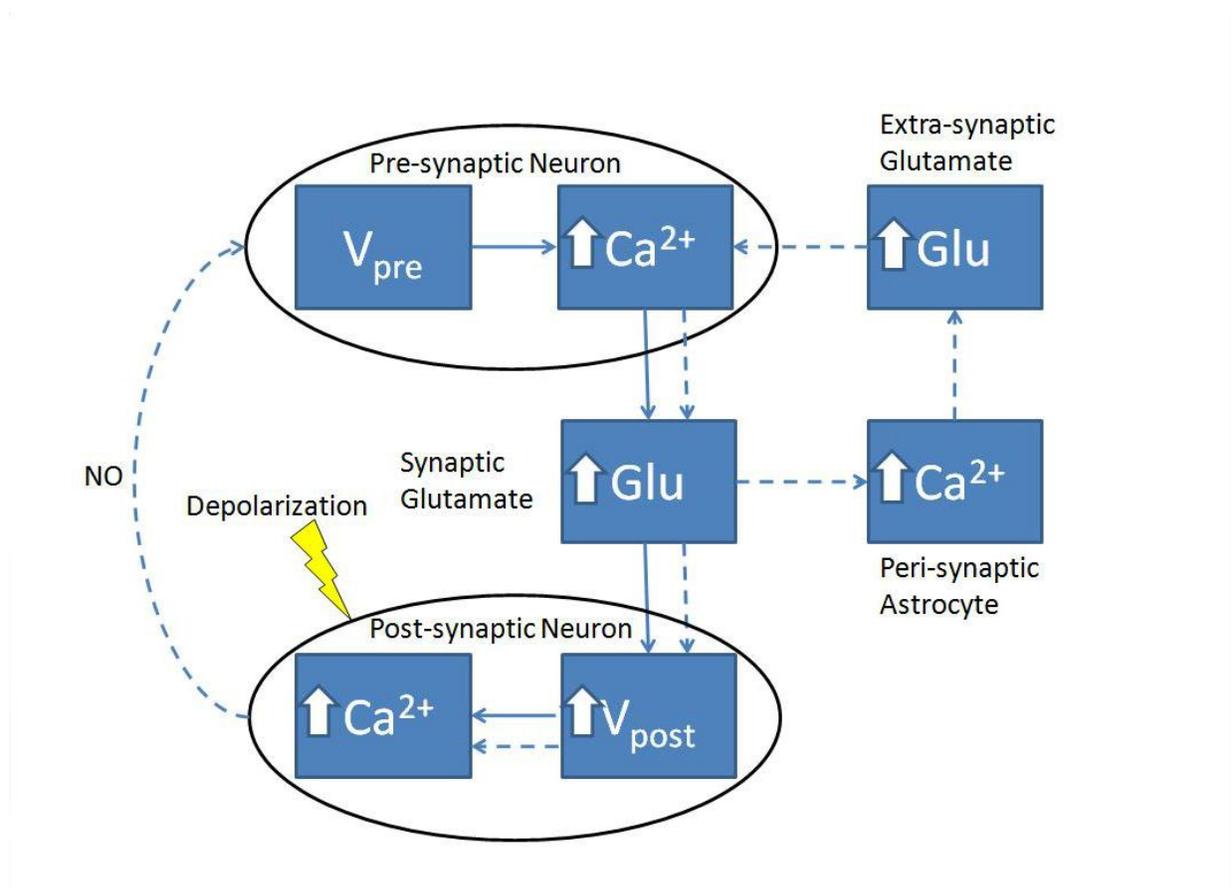

Figure 1. Information flow from pre-synaptic neuron to post-synaptic neuron as modulated by a peri-synaptic astrocyte. Solid line represents astrocyte-independent pathway, while, solid line accompanied by broken line represents astrocyte-dependent pathway. 1) Pre-synaptic AP generated at axon hillock, 2) increased $Ca^{2+}$ concentration in bouton due to opening of VGCCs, 3) elevated synaptic glutamate concentration due to exocytosis of synaptic vesicles – 3a) synaptic glutamate taken up by astrocytic mGluRs instigates the process of astrocyte $Ca^{2+}$ oscillations – 3b) exocytosis of synaptic-like micro-vesicles (SLMVs) leading to an increase in extra-synaptic glutamate concentration; increased glutamate is free to diffuse and binds with mGluRs on the surface of pre-synaptic bouton leading to further flow of $Ca^{2+}$ inside the bouton from intracellular stores responsible for increasing $Ca^{2+}$ concentration transiently, 4) synaptic glutamate is also free to bind with post-synaptic glutamate receptors, (α-amino-3-hydroxy-5-methyl-4-isoxazolepropionic acid receptor) AMPAR here, which leads to an increase in post-synaptic membrane potential, 5) $[Ca^{2+}]$ increase in spine due to fractional $Ca^{2+}$ - current carried by AMPARs and opening of VGCCs during mild-depolarization of post-synaptic neuron. The retrograde signal (NO) denoted by dotted line is to denote its pervasiveness only in an astrocyte-dependent pathway of information processing. The mechanism underlying the proposed role of NO over enhancement of Pr is explained in subsection 2.10.

2.1 Pre-synaptic Action Potential

Action potential (AP) is generated at the axon hillock of the pre-synaptic neuron in response to an applied current density of 10 µA cm$^{-2}$ and frequency 5 Hz. The classical HH (Hodgkin &

Huxley, 1952) paradigm has been used to generate pre-synaptic AP. Since in this paper our focus is not on the detail of the pre-synaptic AP generation, for the sake of simplicity, we have followed the HH model for the pre-synaptic regular spikes generation. Nadkarni & Jung (2003) also followed the HH model for generation of the pre-synaptic AP.

$$C\frac{dV_{pre}}{dt} = I_{app} - g_K n^4 (V_{pre} - V_K) - g_{Na} m^3 h (V_{pre} - V_{Na}) - g_L (V_{pre} - V_L)$$
$$\frac{dx}{dt} = \alpha_x (1-x) - \beta_x x$$
(1)

where $V_{pre}$ is pre-synaptic membrane potential, $I_{app}$ is applied current density, $g_K$, $g_{Na}$ and $g_L$ are potassium, sodium and leak conductance respectively, $V_K$, $V_{Na}$ and $V_L$ are potassium, sodium and leak reversal potential respectively, and $x = m$ (Na$^+$ activation), $h$ (Na$^+$ inactivation) and $n$ (K$^+$ activation). The detail of the HH model can be found in (Hodgkin & Huxley, 1952). The values of the different parameters in equation (1) that have been used in this paper are furnished in the Table 2. $\alpha_x$ and $\beta_x$ for $x = m$, $h$ and $n$ are defined as

$$\alpha_n = \frac{0.01(-V_{pre} - 60)}{\exp(\frac{-V_{pre} - 60}{10}) - 1}, \alpha_m = \frac{0.1(-V_{pre} - 45)}{\exp(\frac{-V_{pre} - 45}{10}) - 1}, \alpha_h = 0.07 \exp(\frac{-V_{pre} - 70}{20}),$$

$$\beta_n = 0.125 \exp(\frac{-V_{pre} - 70}{80}), \beta_m = 4 \exp(\frac{-V_{pre} - 70}{18}), \beta_h = \frac{1}{\exp(\frac{-V_{pre} - 40}{10}) + 1}$$

Table 2: Parameter values used in the HH model (all are from Hodgkin & Huxley, 1952)

| Symbol | Value |
| --- | --- |
| $g_K$ | 36 mS / cm$^2$ |
| $g_{Na}$ | 120 mS / cm$^2$ |
| $g_L$ | 0.3 mS / cm$^2$ |
| $V_K$ | −82 mV |
| $V_{Na}$ | 45 mV |
| $V_L$ | −59.4 mV |

2.2 Pre-synaptic bouton Ca$^{2+}$ dynamics

Input to our mathematical model i.e. the train of AP generated at axon hillock, travels all the way down to the pre-synaptic axon end-feet (or bouton). Its invasion leads to opening of the voltage-gated Ca$^{2+}$ channels (VGCCs), in particular the N-type Ca$^{2+}$ channels, which have been shown to carry majority of the Ca$^{2+}$ required for neurotransmitter release in immature neurons (Mazzanti

& Haydon, 2003; Weber et al., 2010). The total $Ca^{2+}$ concentration, $c_i$, inside the bouton can be divided into two components:

i) $Ca^{2+}$ concentration due to AP, denoted as $c_{fast}$, and

ii) $Ca^{2+}$ concentration due to intracellular stores, $c_{slow}$.

Due to its rapid kinetics, the $Ca^{2+}$ from the VGCCs is termed $c_{fast}$. Similarly, $Ca^{2+}$ from the endoplasmic reticulum (ER) or the intracellular stores is termed $c_{slow}$ due to its slow kinetics. Hence, $c_i$ satisfies the following simple equation

$$c_i = c_{fast} + c_{slow} \Rightarrow \frac{dc_i}{dt} = \frac{dc_{fast}}{dt} + \frac{dc_{slow}}{dt} \qquad (2)$$

The expression for $dc_{fast}/dt$ and $dc_{slow}/dt$ are given by equation (3) and equation (5) respectively. The sensitivity of the $Ca^{2+}$ - sensor of a synaptic vesicle has been studied in detail at the giant Calyx of Held synapses (Schneggenburger & Neher, 2000; Bollman et al. 2000). It was reported that an intracellular $Ca^{2+}$ concentration of ~10 µM is sufficient for neurotransmitter release (Schneggenburger & Neher, 2000; Bollman et al. 2000) in contrast to earlier estimate of ~100 µM (Neher, 1998). The equation governing the $c_{fast}$ consists of simple construction-destruction type formulism and is as follows (Keener & Sneyd, 2009)

$$\frac{dc_{fast}}{dt} = \underbrace{-\frac{I_{Ca} \cdot A_{btn}}{z_{Ca} F V_{btn}} + J_{PMleak}}_{construction} \underbrace{-\frac{I_{PMCa} \cdot A_{btn}}{z_{Ca} F V_{btn}}}_{destruction} \qquad (3)$$

Here, $I_{Ca}$ is the $Ca^{2+}$ current through the N-type channels, $A_{btn}$ is the surface area of the bouton (calculated assuming the bouton to be spherical and using the average volume of a bouton at SC – CA1 synapse due to Koester & Sakmann, 2000), $z_{Ca}$ is the $Ca^{2+}$ ion valence, $F$ is the Faraday's constant, $V_{btn}$ is the volume of the bouton. $I_{PMCa}$ represents the current due to electrogenic plasma-membrane $Ca^{2+}$ ATPase. This high affinity and low capacity pump has known capability to extrude excess of $Ca^{2+}$ out of the cell and it has also been shown that it regulates excitatory synaptic transmission at SC–CA1 synapses (Jensen et al., 2007). To keep it simple, the basic Michaelis-Menton (MM) type formulism has been used for the $Ca^{2+}$ efflux through it (Erler et al., 2004; Blackwell, 2005). $J_{PMleak}$ is the positive leak from the extracellular space into bouton, which makes sure that MM pump does not decrease cytosolic $Ca^{2+}$ to 0 (Blackwell, 2005).

The $Ca^{2+}$ current through VGCCs has been assumed to obey the Ohm's law. The formulation of $I_{Ca}$ follows the single protein level formulation, which is described elsewhere (Erler et al., 2004)

$$I_{Ca} = \rho_{Ca} m_{Ca}^2 \underbrace{g_{Ca}(V_{pre}(t) - V_{Ca})}_{\text{Single open channel}}$$

Here, $\rho_{Ca}$ is the N-type channel protein density which determines the number of Ca$^{2+}$ channels on the membrane of the bouton, $g_{Ca}$ is the single N-type channel conductance, $V_{Ca}$ is the reversal potential of Ca$^{2+}$ ion determined by the Nernst equation (Keener & Sneyd, 2009)

$$V_{Ca} = \frac{RT}{z_{Ca} F} \ln\left(\frac{c_{ext}}{c_i^{rest}}\right), \tag{4}$$

where $R$ is the real gas constant, $T$ is the absolute temperature, $c_{ext}$ is the extracellular Ca$^{2+}$ concentration, $c_i^{rest}$ is the total intracellular Ca$^{2+}$ concentration at rest. It is assumed that a single N-type channel consists of two-gates. $m_{Ca}$ denotes the opening probability of a single gate. A single N-type channel is open only when both the gates are open. Hence, $m_{Ca}^2$ is the single channel open probability. The time dependence of the single channel open probability is governed by an HH-type formulation

$$\frac{dm_{Ca}}{dt} = \frac{(m_{Ca}^\infty - m_{Ca})}{\tau_{m_{Ca}}},$$

where $m_{Ca}^\infty$ is the Boltzmann-function fitted by Ishikawa et al. (2005) to the whole cell current of an N-type channel, $m_{Ca}$ approaches its asymptotic value $m_{Ca}^\infty$ with a time constant $\tau_{m_{Ca}}$. The mathematical expression of other parameters used in equation (3) is as follows

$$I_{PMCa} = v_{PMCa} \frac{c_i^2}{c_i^2 + K_{PMCa}^2}, \quad J_{PMleak} = v_{leak}(c_{ext} - c_i), \quad m_{Ca}^\infty = \frac{1}{1 + \exp\left((V_{m_{Ca}} - V_{pre})/k_{m_{Ca}}\right)}$$

Here, $v_{PMCa}$ is the maximum PMCa current density, determined through computer simulations, so that $c_i$ is maintained at its resting concentration. All other parameter values used for simulation are listed in Table 3.

Table 3: Parameters used for Bouton Ca$^{2+}$ dynamics

| Symbol | Description | Value | Reference |
|---|---|---|---|
| $F$ | Faraday's constant | 96487 C / mole | |
| $R$ | Real gas constant | 8.314 J / K | |
| $T$ | Absolute Temperature | 293.15 K | Temperature in Perea & Araque (2007) |
| $z_{Ca}$ | Calcium valence | 2 | |
| $A_{btn}$ | Surface area of bouton | 1.24 μm$^2$ | Koester & Sakmann, 2000 |

| $V_{btn}$ | Volume of bouton | 0.13 μm³ | Koester & Sakmann, 2000 |
|---|---|---|---|
| $\rho_{Ca}$ | N-type channel density | 3.2 / μm² | This paper† |
| $g_{Ca}$ | N-type channel conductance | 2.3 pS | Weber et al. 2010 |
| $V_{Ca}$ | Reversal potential of $Ca^{2+}$ ion | 125 mV | Calculated using equation (4) |
| $v_{PMCa}$ | Maximum PMCa current | 0.4 μA / cm² | See text; Page No. 7 |
| $K_{PMCa}$ | $Ca^{2+}$ concentration at which $v_{PMCa}$ is halved | 0.1 μM | Erler et al. 2004 |
| $v_{leak}$ | Maximum leak of $Ca^{2+}$ through plasma membrane | 2.66 x 10⁻⁶ / ms | See text; Page No. 6 |
| $c_i^{rest}$ | Resting Intracellular $Ca^{2+}$ concentration | 0.1 μM | |
| $c_{ext}$ | External $Ca^{2+}$ concentration | 2 mM | Extracellular [$Ca^{2+}$] in Perea & Araque (2007) |
| $V_{mCa}$ | Half-activation voltage of N-type $Ca^{2+}$ channel | -17 mV | Ishikawa et al. 2005 |
| $k_{mCa}$ | Slope factor of N-type channel activation | 8.4 mV | Ishikawa et al. 2005 |
| $\tau_{m_{Ca}}$ | N-type channel time constant | 10 ms | Ishikawa et al. 2005 |
| $c_1$ | Ratio of ER volume to volume of Bouton | 0.185 | Shuai & Jung, 2002 |
| $v_1$ | Maximum IP₃ receptor flux | 30 / s | See text; Page No. 9 |
| $v_2$ | $Ca^{2+}$ leak rate constant | 0.2374 / s | See text; Page No. 9 |
| $v_3$ | SERCA maximal pump rate | 90 μM / s | See text; Page No. 9 |
| $k_3$ | SERCA dissociation constant | 0.1 μM | Jafri & Keizer, 1995 |
| $d_1$ | IP₃ dissociation constant | 0.13 μM | Shuai & Jung, 2002 |
| $d_2$ | Inhibitory $Ca^{2+}$ dissociation constant | 1.049 μM | Shuai & Jung, 2002 |
| $d_3$ | IP₃ dissociation constant | 943.4 nM | Shuai & Jung, 2002 |
| $d_5$ | Activation $Ca^{2+}$ dissociation constant | 82.34 nM | Shuai & Jung, 2002 |
| $a_2$ | Inhibitory $Ca^{2+}$ binding constant | 0.2 μM / s | Shuai & Jung, 2002 |
| $v_g$ | Maximum production rate of IP₃ | 0.062 μM / s | Nadkarni & Jung, 2008 |
| $k_g$ | Glutamate concentration at which $v_g$ is halved | 0.78 μM | Nadkarni & Jung, 2008 |
| $\tau_p$ | IP₃ degradation constant | 0.14 / s | Nadkarni & Jung, 2008 |
| $p_0$ | Initial IP₃ concentration | 0.16 μM | Nadkarni & Jung, 2008 |

† Determined through computer simulations so that the average Pr lies between 0.2–0.3 (when astrocyte is not stimulated) similar to the experiments of Perea & Araque (2007).

Although, the flux of $Ca^{2+}$ from the endoplasmic reticulum (ER) is mainly controlled by two types of receptors (or $Ca^{2+}$ channels) i) the inositol (1,4,5)-trisphosphate receptor (IP₃R) and ii) the ryanodine receptor (RyR) (Sneyd & Falcke, 2005). We have assumed ER to have IP₃R alone, because it has been shown to modulate ganglionic LTP (Vargas et al., 2010). The inositol (1,4,5)-triphosphate (IP₃) is necessary for the release of $Ca^{2+}$ through IP₃R (on the surface of the ER). It is produced when the glutamate (agonist) binds with the mGluR (receptor) and causes via G-protein link to phospholipase C (PLC), the cleavage of phosphotidylinositol (4,5)- bisphosphate (PIP$_2$) to produce IP$_3$ and diacylglycerol (DAG). We have used the conventional Li-Rinzel model (L-R model) (Li & Rinzel, 1994) to formulate this slower Ca$^{2+}$ signaling process. A few modifications were made to the L-R model. The L-R model assumes that, total intracellular concentration, $c_0$, is conserved and determines the ER Ca$^{2+}$ concentration, $c_{ER}$, using the following relation

$$c_{ER} = \frac{(c_0 - c_i)}{c_1}.$$

Such an assumption is not valid in the present model because of the presence of membrane fluxes, namely, $I_{Ca}$ and $I_{PMCa}$. Also, in the L-R model intracellular IP$_3$ concentration is used as a control parameter. This is not supposed to hold in an active TpS. Hence, two additional equations governing ER Ca$^{2+}$ concentration and IP$_3$ concentration have been incorporated in the classical L-R model. The IP$_3$ production term was made glutamate dependent to study the effect of astrocytic Ca$^{2+}$ over $c_i$. The mathematical model governing the $c_{slow}$ dynamics is as follows

$$\begin{aligned}
\frac{dc_{slow}}{dt} &= -c_1 v_1 m_\infty^3 n_\infty^3 q^3 (c_i - c_{ER}) - \frac{v_3 c_i^2}{k_3^2 + c_i^2} - c_1 v_2 (c_i - c_{ER}), \\
\frac{dc_{ER}}{dt} &= -\frac{1}{c_1} \frac{dc_{slow}}{dt}, \\
\frac{dp}{dt} &= v_g \frac{g_a^{0.7}}{k_g^{0.7} + g_a^{0.7}} - \tau_p (p - p_0), \\
\frac{dq}{dt} &= \alpha_q (1-q) - \beta_q q.
\end{aligned} \quad (5)$$

With, $m_\infty = \frac{p}{p + d_1}$, $n_\infty = \frac{c_i}{c_i + d_5}$, $\alpha_q = a_2 d_2 \frac{p + d_1}{p + d_3}$, $\beta_q = a_2 c_i$. In equation (5), the first term on the right hand side of $dc_{slow}/dt$ denotes Ca$^{2+}$ flux from ER to the intracellular space through IP$_3$R, second term is the Ca$^{2+}$ flux pumped from the intracellular space into ER, third term is the leak of Ca$^{2+}$ ions from ER to intracellular space. $c_{ER}$ is the ER Ca$^{2+}$ concentration, $c_1$ is the ratio of volume of ER to volume of bouton, $v_1$ is maximal IP$_3$R flux rate, $v_2$ is the maximal Ca$^{2+}$ leak from ER to intracellular space, $v_3$ maximal Sarco-Endoplasmic Reticulum ATPase (SERCA) pump rate, $p$ is the intracellular IP$_3$ concentration, $g_a$ is the glutamate concentration in the extra-synaptic cleft, $q$ is the fraction of activated IP$_3$R.

Most of the values of $v_1$, $v_2$, $v_3$ mentioned in literature are for closed-cell dynamics which is not the case here. Hence, the values of $v_1$, $v_2$, $v_3$ were fixed through computer simulation runs so that Ca$^{2+}$ homeostasis is maintained inside the bouton and its organelles with or without pre-synaptic stimulus. Details of parameters are as listed in Table 3.

## 2.3 Glutamate release dynamics in bouton

The sensitivity of the $Ca^{2+}$ - sensor is a critical parameter which decides the release of glutamate from the synaptic vesicles. The study of $Ca^{2+}$ sensitivity of the $Ca^{2+}$ - sensor at hippocampal synapses is hampered due to its small size (Wang et al., 2009). However, $Ca^{2+}$ sensitivity of the $Ca^{2+}$ - sensor has been studied in detail at the giant Calyx of Held synapses (Schneggenburger & Neher, 2000; Bollman et al., 2000) and squid giant synapse (Llinas, 1999). Schneggenburger & Neher (2000) suggested that an intracellular $Ca^{2+}$ concentration of ~10 µM is sufficient for the neurotransmitter release at the giant Calyx of Held synapses. This estimate is 10 folds smaller than our estimate for hippocampal synapses i.e. 100 µM (Neher, 1998; Nadkarni & Jung, 2008). The tour-de-force attempt (Schneggenburger & Neher, 2000; Bollman et al., 2000) at the Calyx of Held is not easy to implement at small nerve terminals of hippocampal synapses. We have assumed the sensitivity of the $Ca^{2+}$ - sensor to be the same as that of the Calyx of Held. The kinetic model governing the $Ca^{2+}$ binding to the $Ca^{2+}$ sensor is given by the following equations (Bollman et al., 2000)

$$X \underset{\beta}{\overset{5\alpha c_i}{\rightleftharpoons}} X(c_i)_1 \underset{2\beta}{\overset{4\alpha c_i}{\rightleftharpoons}} X(c_i)_2 \underset{3\beta}{\overset{3\alpha c_i}{\rightleftharpoons}} X(c_i)_3 \underset{4\beta}{\overset{2\alpha c_i}{\rightleftharpoons}} X(c_i)_4 \underset{5\beta}{\overset{\alpha c_i}{\rightleftharpoons}} X(c_i)_5 \underset{\delta}{\overset{\gamma}{\rightleftharpoons}} X(c_i)_5^*, \qquad (6)$$

where, α and β are the $Ca^{2+}$ association and dissociation rate constants respectively, γ and δ are $Ca^{2+}$ independent isomerisation constants. $X$ is the $Ca^{2+}$ sensor, of a synaptic vesicle, with no $Ca^{2+}$ bound, $X(c_i)_1$ is $Ca^{2+}$ sensor with one $Ca^{2+}$ bound, likewise $X(c_i)_5$ is the $Ca^{2+}$ sensor with five $Ca^{2+}$ bound and $X(c_i)_5^*$ is the isomer of $X(c_i)_5$ which is ready for glutamate release. Hippocampal synapses are known as low-fidelity synapses (Nadkarni & Jung, 2008). We have assumed an active zone consisting of two-docked synaptic vesicles (Danbolt, 2001; Nikonenko & Skibo, 2006). Since there are fewer synaptic vesicles, the number of vesicles with 5 $Ca^{2+}$ ions bound cannot be determined by the average of vesicle pool. Therefore the fraction of the docked vesicles ready to be released $f_r$, has been determined using dynamic Monte-Carlo simulations (Fall et al., 2002) of kinetic equation (6) and depends on the $X(c_i)_5^*$ state.

Apart from the evoked release of glutamate, the spontaneous release of glutamate can also occur in our model. It is known that spontaneous release of neurotransmitter depends upon the pre-synaptic $Ca^{2+}$ concentration (Emptage et al., 2001; Nadkarni & Jung, 2008). Also, it is known that the spontaneous release takes place when pre-synaptic $Ca^{2+}$ is low i.e. when pre-synaptic membrane is not depolarized (Atluri & Regehr, 1998; Hagler & Goda, 2001). The fraction of synaptic vesicles ready to be release spontaneously, $f_r$, is assumed to be a Poisson process with the following rate

$$\lambda(c_i) = a_3 \left(1 + \exp\left(\frac{a_1 - c_i}{a_2}\right)\right)^{-1}. \qquad (7)$$

The formulation for the rate of spontaneous release is from Nadkarni & Jung (2008). However, we have to modify the parameter values, in equation (7), because as per their choice of values and system setup, the frequency of spontaneously released vesicles was as high as 19 per sec (determined through simulation runs for over 10000 times). However, the experimentally determined frequency of spontaneous vesicle release in presence of an astrocyte is in between 1 – 3 per sec (Kang et al, 1998). Thus, we determined the values of $a_1$, $a_2$ and $a_3$ so that the frequency of spontaneous vesicle release is between 1 – 3 Hz. The vesicle fusion and recycling process is governed by the Tsodyks & Markram model (Tsodyks & Markram, 1997). A slight modification has been made to the Tsodyks & Markram Model (TMM) to make the vesicle fusion process $f_r$ dependent. The modified TMM is as follows

$$\frac{dR}{dt} = \frac{I}{\tau_{rec}} - f_r \cdot R$$
$$\frac{dE}{dt} = -\frac{E}{\tau_{inact}} + f_r \cdot R \quad (8)$$
$$I = 1 - R - E$$

where $R$ is the fraction of the releasable vesicles inside the bouton, $E$ is the fraction of the effective vesicles in the synaptic cleft and $I$ is the fraction of the inactive vesicles undergoing recycling process, $f_r$ has the value (0, 0.5, 1) provided that the pre-synaptic membrane is depolarized (for evoked vesicle release) or pre-synaptic membrane is not depolarized (for spontaneous vesicle release), corresponding to the number of vesicles ready to be released (0, 1, 2), which is determined by the stochastic simulation of the kinetic model in equation (6) (for evoked vesicle release) or generating a Poisson random variable with the rate given by equation (7) (for spontaneous vesicle release). $\tau_{inact}$ and $\tau_{rec}$ are the time constants of vesicle inactivation and recovery respectively. Once a vesicle is released whether evoked or spontaneous, the vesicle release process remains inactivated for a period of 6.34 ms (Nadkarni & Jung, 2008). The parametric values used for simulation are listed in Table 4.

Table 4: Parameters used for Glutamate dynamics in bouton and cleft

| Symbol | Description | Value | Reference |
| --- | --- | --- | --- |
| α | $Ca^{2+}$ association rate constant | 0.3 / μM ms | Bollman et al. 2000 |
| β | $Ca^{2+}$ dissociation rate constant | 3 / ms | Bollman et al. 2000 |
| γ | Isomerization rate constant (forward) | 30 / ms | Bollman et al. 2000 |
| δ | Isomerization rate constant (backward) | 8 / ms | Bollman et al. 2000 |
| $\tau_{rec}$ | Vesicle recovery time constant | 800 ms | Tsodyks & Markram, 1997 |
| $\tau_{inac}$ | Vesicle inactivation time constant | 3 ms | Tsodyks & Markram, 1997 |
| $a_1$ | $Ca^{2+}$ concentration at which λ is halved | 50 μM | See text; Page No. 11 |

| $a_2$ | Slope factor of spontaneous release rate λ | 5 μM | See text; Page No. 11 |
|---|---|---|---|
| $a_3$ | Maximum spontaneous release rate | 0.85 / ms | See text; Page No. 11 |
| $n_v$ | Number of docked vesicle | 2 | Nikonenko & Skibo, 2006 |
| $g_v$ | Glutamate concentration in single vesicle | 60 mM | Montana et al., 2006 |
| $g_c$ | Glutamate clearance rate constant | 10 / ms | Destexhe et al., 1998 |

2.4 Glutamate dynamics in synaptic cleft

The effective concentration of glutamate in the synaptic cleft is decisive in controlling synaptic efficacy unless the post-synaptic receptors are already saturated which is possible in response to a high frequency stimulus (HFS). In our model, the glutamate available in the synaptic cleft was made available to the post-synaptic receptors and mGluR on the surface of astrocyte. Since E gives the effective fraction of the vesicles in the synaptic cleft, the estimated glutamate concentration in synaptic cleft can be represented mathematically as

$$\frac{dg}{dt} = n_v \cdot g_v \cdot E - g_c \cdot g \quad (9)$$

Here $g$ is the glutamate concentration in the synaptic cleft, $n_v$ is the number of docked vesicle, $g_v$ is the vesicular glutamate concentration and $g_c$ is the rate of glutamate clearance i.e., re-uptake by neuron or astrocyte. Using this simple dynamics we could achieve the estimated range of glutamate concentration 0.24 - 11 mM in synaptic cleft (Danbolt, 2001; Franks et al., 2002) and time course of glutamate in synaptic cleft 2 ms (Franks et al., 2002; Nadkarni & Jung, 2007). Although a similar equation can be used to model the glutamate dynamics in other synapses, although, one may have to use different constant values. Thus the present formulation can be considered specific to the SC – CA1 synapse.

2.5 Astrocyte $Ca^{2+}$ dynamics

Porter & McCarthy (1996) demonstrated that the glutamate released from the SC leads to an increase in astrocytic $Ca^{2+}$ via an mGluR-dependent pathway *in situ*. Also recently De Pitta et al. (2009) proposed a mathematical model, termed as G-ChI, for astrocytic $Ca^{2+}$ concentration increase through an mGluR-dependent pathway. They used glutamate as a stimulating parameter to evoke increases in astrocytic $Ca^{2+}$ concentration, while in our model glutamate comes through SC in an activity-dependent manner (see equation (9)). The G-ChI model uses the conventional L-R model for astrocytic $Ca^{2+}$ concentration '$c_a$' with some specific terms for the astrocytic $IP_3$ concentration '$p_a$.' The model incorporates PLCβ and PLCδ dependent $IP_3$ production. It also incorporates inositol polyphosphate 5-phosphatase (IP-5P) and $IP_3$ 3-kinase ($IP_3$-3K) dependent $IP_3$ degradation (for details see De Pitta et al, 2009). It is a very detailed, astrocyte specific, model which exhibits $IP_3$ oscillations apart from $Ca^{2+}$ oscillations. However the exact significance of IP$_3$ oscillation is not yet known (De Pitta et al, 2009). The G-ChI model is a closed-cell model (Keener & Sneyd, 2009) i.e. without membrane fluxes. In such models $c_a$ primarily depends upon two parameters namely, i) flux from ER into cytosol and ii) the maximal pumping capacity of the SERCA pump. IP$_3$R is found in clusters of variable size (Nadkarni & Jung, 2007). The exact size of cluster is supposed to vary from cell to cell but following Nadkarni & Jung (2007) we have assumed it be 20. Considering such a small size of IP$_3$R cluster stochasticity does play a significant role (see Shuai & Jung, 2002), and hence instead of using an ordinary differential equation for gating parameter $h_a$, we have used Langevin equation (Fox, 1997; Shuai & Jung, 2002). The model can be represented as follows

$$\frac{dc_a}{dt} = -r_{c_a} m_{\infty,a}^3 n_{\infty,a}^3 h_a^3 \left(c_0 - \left(1 + c_{1,a}\right)c_a\right) - v_{ER} \frac{c_a^2}{c_a^2 + K_{ER}^2} - r_L \left(c_0 - \left(1 + c_{1,a}\right)c_a\right) \tag{10}$$

$$\frac{dp_a}{dt} = v_\beta \cdot \text{Hill}\left(g^{0.7}, K_R \left(1 + \frac{K_p}{K_R} \text{Hill}(C, K_\pi)\right)\right) + \frac{v_\delta}{1 + \frac{p_a}{k_\delta}} \text{Hill}\left(c_a^2, K_{PLC\delta}\right)$$
$$+ v_{3K} \text{Hill}\left(c_a^4, K_D\right) \text{Hill}\left(p_a, K_3\right) - r_{5p_a} p_a \tag{11}$$

$$\frac{dh_a}{dt} = \alpha_{h_a}\left(1 - h_a\right) - \beta_{h_a} h_a + G_h(t) \tag{12}$$

The first term on the right hand side of equation (10) is due to efflux of Ca$^{2+}$ from ER to intracellular space, the second term represents removal of excess Ca$^{2+}$ from intracellular space in to ER by the SERCA pump, the third term is a constant leak of Ca$^{2+}$ concentration from ER to intracellular space. $r_{c_a}$ is the maximal rate of Ca$^{2+}$ flux from ER, $m_{\infty,a}^3 \cdot n_{\infty,a}^3 \cdot h_a^3$ governs the opening probability of the IP$_3$R cluster, $c_0$ is the cell-averaged total of the free Ca$^{2+}$ concentration, $c_{1,a}$ is the ratio of ER and cytosol volume, $v_{ER}$ is the maximal rate of Ca$^{2+}$ uptake by the SERCA pump, $K_{ER}$ is the SERCA pump Ca$^{2+}$ affinity, $r_L$ is the maximal rate of Ca$^{2+}$ leakage from the ER. The first two terms on the right hand side of equation (11) are IP$_3$ producing terms due to PLC**β** and PLC**δ** respectively; the last two terms are IP$_3$ degradation terms due to IP$_3$-3K and IP-5P respectively (for a systematic derivation, see De Pitta et al (2009)). Equation (12) is analogous to gating variable $q$ in equation (5) except for $G_h(t)$, which is a zero mean, uncorrelated, Gaussian white-noise term with co-variance function (Nadkarni & Jung, 2007)

$$\left\langle G_h(t) G_h(t')\right\rangle = \frac{\alpha_{h_a}(1 - h_a) + \beta_{h_a} h_a}{N_{IP_3}} \delta(t - t')$$

Here, $\delta(t)$ is the Dirac-delta function, $t$ and $t'$ are time and $\dfrac{\alpha_{h_a}(1-h_a)+\beta_{h_a}h_a}{N_{IP_3}}$ is the spectral density (Coffey et al. 2005). Mathematical expression of all other parameters in equations (10) - (12) are

$$m_{\infty,a} = \text{Hill}(p_a, d_1),\ n_{\infty,a} = \text{Hill}(c_a, d_5),\ \text{Hill}(x^n, K) = \dfrac{x^n}{x^n + K^n},$$

$$\alpha_{h_a} = a_2 d_2 \dfrac{p_a + d_1}{p_a + d_3},\ \beta_{h_a} = a_2 c_a$$

$\text{Hill}(x^n, K)$ is the generic Hill function (De Pitta et al., 2009). Hill function is generally used to model those reactions whose reaction velocity curve is not hyperbolic (Keener & Sneyd, 2009). Details of all parameters are as listed in Table 5.

Table 5: Parameters used for astrocyte $Ca^{2+}$ dynamics

| Symbol | Description | Value | Reference |
|---|---|---|---|
| $r_{c_a}$ | Maximal IP$_3$R flux | 6 / s | De Pitta et al. 2009 |
| $r_L$ | Maximal rate of Ca$^{2+}$ leak from ER | 0.11 / s | De Pitta et al. 2009 |
| $c_0$ | Total cell free Ca$^{2+}$ concentration | 2 μM | De Pitta et al. 2009 |
| $c_{1,a}$ | Ratio of ER volume to cytosol volume | 0.185 | De Pitta et al. 2009 |
| $v_{ER}$ | Maximal rate of SERCA uptake | 0.9 μM / s | De Pitta et al. 2009 |
| $K_{ER}$ | SERCA Ca$^{2+}$ affinity | 0.1 μM | De Pitta et al. 2009 |
| $d_1$ | IP$_3$ dissociation constant | 0.13 μM | De Pitta et al. 2009 |
| $d_2$ | Ca$^{2+}$ inactivation dissociation constant | 1.049 μM | De Pitta et al. 2009 |
| $d_3$ | IP$_3$ dissociation constant | 0.9434 μM | De Pitta et al. 2009 |
| $d_5$ | Ca$^{2+}$ activation dissociation constant | 0.08234 μM | De Pitta et al. 2009 |
| $a_2$ | IP$_3$R binding rate for Ca$^{2+}$ Inhibition | 2 / s | De Pitta et al. 2009 |
| $N$ | Number of IP$_3$R in a cluster | 20 | Nadkarni & Jung, 2007 |
| Glutamate-dependent IP$_3$ production | | | |
| $v_\beta$ | Maximal rate of IP$_3$ production by PLCβ | 0.5 μM / s | De Pitta et al. 2009 |
| $K_R$ | Glutamate affinity of the receptor | 1.3 μM | De Pitta et al. 2009 |
| $K_p$ | Ca$^{2+}$/PKC-dependent inhibition factor | 10 μM | De Pitta et al. 2009 |
| $K_\pi$ | Ca$^{2+}$ affinity of PKC | 0.6 μM | De Pitta et al. 2009 |
| Glutamate-independent IP$_3$ production | | | |
| $v_\delta$ | Maximal rate of IP$_3$ production by PLCδ | 0.05 μM / s | De Pitta et al. 2009 |
| $K_{PLC\delta}$ | Ca$^{2+}$ affinity of PLCδ | 0.1 μM | De Pitta et al. 2009 |

| $k_\delta$ | Inhibition constant of PLCδ activity | 1.5 μM | De Pitta et al. 2009 |
|---|---|---|---|
| IP$_3$ degradation | | | |
| $r_{5pa}$ | Maximal rate of degradation by IP-5P | 0.05 / s | De Pitta et al. 2009 |
| $v_{3K}$ | Maximal rate of degradation by IP$_3$-3K | 2 μM / s | De Pitta et al. 2009 |
| $K_D$ | Ca$^{2+}$ affinity of IP$_3$-3K | 0.7 μM | De Pitta et al. 2009 |
| $K_3$ | IP$_3$ affinity of IP$_3$-3K | 1 μM | De Pitta et al. 2009 |

2.6 Gliotransmitter release dynamics in astrocyte

Astrocytes in vitro and in vivo have been shown to contain synaptic-like micro-vesicles (SLMVs) (Bezzi et al., 2004; Marchaland et al., 2008) and express the protein machinery necessary for Ca$^{2+}$-dependent exocytosis process (Bezzi et al., 2004; Montana et al., 2006; Verkhratsky & Butt, 2007; Parpura & Zorec, 2010). However, the exact mechanism by which astrocyte release gliotransmitter is yet to be determined (Wenker, 2010; Ben Achour et al., 2010). Parpura & Haydon (2000) determined Ca$^{2+}$ dependency of glutamate release from the hippocampal astrocytes. In their study the Hill coefficient for glutamate release was 2.1 – 2.7 suggesting at least two Ca$^{2+}$ ions are must for a possible gliotransmitter release (Parpura & Haydon, 2000). Following the experimental observation of Parpura & Haydon (2000), in this manuscript, it has been assumed that the binding of three Ca$^{2+}$ ions to an SLMV makes it fusogenic (ready to be released). The percent of fusogenic SLMVs in response to a mechanical stimulation and the size of readily releasable pool of SLMVs in the astrocytes have been determined recently (Malarkey & Parpura, 2011). It is assumed that the gliotransmitter release site contains three independent gates ($S_1 – S_3$), for the three Ca$^{2+}$ ions to bind, with different opening and closing constants. The model governing the gliotransmitter release site activation is based on Bertram et al. (1996) and is as follows

$$c_a + C_j \underset{k_j^-}{\overset{k_j^+}{\rightleftarrows}} O_j \qquad j = 1, 2, 3$$

Where, $k_j^+$ and $k_j^-$ are the opening and closing rates of the gate '$S_j$'; $C_j$ and $O_j$ are the closing and opening probability of the gate $S_j$. The temporal evolution of the open gate '$O_j$' can be expressed as

$$\frac{dO_j}{dt} = k_j^+ \cdot c_a - \left(k_j^+ \cdot c_a + k_j^-\right) \cdot O_j \tag{13}$$

An SLMV is ready to be released when all three gates are bound with Ca$^{2+}$. Hence, the probability that an SLMV is ready to be released is given by

$$f_r^a = O_1 \cdot O_2 \cdot O_3. \tag{14}$$

The dissociation constants of gates $S_1 - S_3$ are 108 nM, 400 nM, and 800 nM. The time constants for gate closure $\left(1/k_j^-\right)$ are 2.5 s, 1s, and 100 ms, respectively. The dissociation constants and time constants for $S_1$ and $S_2$ are same as in Bertram et al (1996). While the dissociation constant and time constant for gate $S_3$ were fixed through computer simulations to fit the experimentally determined probability of fusogenic SLMVs found recently (Malarkey & Parpura, 2011). Similar to pre-synaptic bouton, the SLMV fusion and recycling process is modeled using TMM after making the release process $f_r^a$ dependent. The governing model is as follows

$$\frac{dR_a}{dt} = \frac{I_a}{\tau_{rec}^a} - \Theta\left(c_a - c_a^{thresh}\right) \cdot f_r^a \cdot R_a$$
$$\frac{dE_a}{dt} = -\frac{E_a}{\tau_{inact}^a} + \Theta\left(c_a - c_a^{thresh}\right) \cdot f_r^a \cdot R_a \quad (15)$$
$$I_a = 1 - R_a - E_a$$

Here $R_a$ is the fraction of the readily releasable SLMV inside the astrocyte, $E_a$ is the fraction of the effective SLMV in the extra-synaptic cleft and $I_a$ is the fraction of the inactive SLMV undergoing endocytosis or re-acidification process. $\Theta$ is the Heaviside function and $c_a^{thresh}$ is the threshold of astrocyte $Ca^{2+}$ concentration necessary for release site activation (Parpura & Haydon, 2000). $\tau_{inact}^a$ and $\tau_{rec}^a$ are the time constants of inactivation and recovery respectively.

Table 6: Parameters used for Glutamate dynamics in astrocyte and extra-synaptic cleft

| Symbol | Description | Value | Reference |
| --- | --- | --- | --- |
| $k_1^+$ | $Ca^{2+}$ association rate for $S_1$ | 3.75 x $10^{-3}$/ μM ms | Bertram et al. 1996 |
| $k_1^-$ | $Ca^{2+}$ dissociation rate for $S_1$ | 4 x $10^{-4}$ / ms | Bertram et al. 1996 |
| $k_2^+$ | $Ca^{2+}$ association rate for $S_2$ | 2.5 x $10^{-3}$/μM ms | Bertram et al. 1996 |
| $k_2^-$ | $Ca^{2+}$ dissociation rate for $S_2$ | 1 x $10^{-3}$ / ms | Bertram et al. 1996 |
| $k_3^+$ | $Ca^{2+}$ association rate for $S_3$ | 1.25 x $10^{-2}$/ μM ms | See text; Page No. 15 |
| $k_3^-$ | $Ca^{2+}$ dissociation rate for $S_3$ | 1 x $10^{-3}$ / ms | See text; Page No. 15 |
| $\tau_{rec}^a$ | Vesicle recovery time constant | 800 ms | Tsodyks & Markram, 1997 |
| $\tau_{inac}^a$ | Vesicle inactivation time constant | 3 ms | Tsodyks & Markram, 1997 |
| $c_a^{thresh}$ | Astrocyte response threshold | 196.69 nM | Parpura & Haydon, 2000 |
| $n_a^v$ | SLMV ready to be released | 12 | Malarkey & Parpura, 2011 |
| $g_a^v$ | Glutamate concentration in one SLMV | 20 mM | Montana et al. 2006 |
| $g_a^c$ | Glutamate clearance rate from the extra-synaptic cleft | 10 / ms | Destexhe et al. 1998 |

2.7 Glutamate dynamics in extra-synaptic cleft

The glutamate in the extra-synaptic cleft $g_a$, has been modeled in a similar way to equation (9). This glutamate acts on extra-synaptically located group I mGluR of the pre-synaptic bouton. It is used as an input in the IP$_3$ production term of equation (5). The SLMVs of the astrocytes are not as tightly packed as in neurons (Bezzi et al., 2004). Thus it is assumed that each SLMV contains 20 mM of glutamate (Montana et al., 2006). The mathematical equation governing the glutamate dynamics are as follows

$$\frac{dg_a}{dt} = n_a^v \cdot g_a^v \cdot E_a - g_a^c \cdot g_a, \qquad (16)$$

where $g_a$ is the glutamate concentration in the extra-synaptic cleft, $n_a^v$ represents the size of readily releasable pool of SLMVs, $g_a^v$ is the glutamate concentration in one SLMV, $g_a^c$ is the clearance rate of glutamate from the cleft due to diffusion and/or re-uptake.

2.8 Post-synaptic membrane potential

The dendritic spine-head is assumed to be of mushroom type. Its volume is taken to be 0.9048 μm$^3$ (assuming a spherical spine-head of radius 0.6 μm (Dumitriu et al, 2010)). The specific capacitance and specific resistance of the spine-head is assumed to be 1 μF / cm$^2$ and 10000 Ω cm$^2$, respectively (Koch, 1999). The spine-neck is assumed to be of negligible diameter. Given the size of the spine we can calculate its actual resistance as

$$R_m = \frac{R_M}{A_{spine}} \qquad (17)$$

Where, $R_m$ is actual resistance, $R_M$ is specific resistance and $A_{spine}$ is the area of spine-head. NMDAR and AMPAR are co-localized in most of the glutamatergic synapses, most of which are found in dendrite spines (Franks et al., 2002). Since we were modeling NMDAR-independent LTP hence NMDAR was not mathematically modeled. The post-synaptic potential change has been modeled using a passive membrane mechanism (Tsodyks & Markram, 1997)

$$\tau_{post} \frac{dV_{post}}{dt} = -(V_{post} - V_{post}^{rest}) - R_m \cdot (I_{soma} + I_{AMPA}), \qquad (18)$$

where $\tau_{post}$ is the post-synaptic membrane time constant, $V_{post}^{rest}$ is the post-synaptic resting membrane potential, $I_{soma}$ is the current injected in soma to raise the post-synaptic potential to -30 mV, $I_{AMPA}$ is the AMPAR current and is given by the following expression

$$I_{AMPA} = g_{AMPA} m_{AMPA} (V_{post} - V_{AMPA}) \qquad (19)$$

where $g_{AMPA}$ is the conductance of the AMPARs at single synapses, $V_{AMPA}$ is the reversal potential of the AMPAR and $m_{AMPA}$ is the gating variable of AMPAR. Although there exist more comprehensive and detailed models for AMPAR gating (Destexhe et al., 1998) we have used a simple 2-state model to avoid redundant complexity. The simple 2-state model is known to retain most of the important qualitative properties of AMPAR (Destexhe et al., 1998). AMPAR gating is governed by the following HH-type formulism (Destexhe et al., 1998)

$$\frac{dm_{AMPA}}{dt} = \alpha_{AMPA} g \left(1 - m_{AMPA}\right) - \beta_{AMPA} m_{AMPA}$$

Here $\alpha_{AMPA}$ is the opening rate of the receptor, $\beta_{AMPA}$ is the closing rate of the receptor and $g$ is the glutamate concentration in the cleft given by equation (9). The parameter values are as listed in Table 7.

Table 7: List of parameters used for post-synaptic potential generation

| Symbol | Description | Value | Reference |
| --- | --- | --- | --- |
| $R_m$ | Actual resistance of the spine-head | $0.7985 \times 10^5$ M$\Omega$ | Calculated using equation (17) |
| $V_{post}^{rest}$ | Post-synaptic resting membrane potential | -70 mV | |
| $\tau_{post}$ | Post-synaptic membrane time constant | 50 ms | Tsodyks & Markram, 1997 |
| $g_{AMPA}$ | AMPAR conductance | 0.35 nS | Destexhe et al. 1998 |
| $V_{AMPA}$ | AMPAR reversal potential | 0 mV | Destexhe et al. 1998 |
| $\alpha_{AMPA}$ | AMPAR forward rate constant | 1.1 μM / s | Destexhe et al. 1998 |
| $\beta_{AMPA}$ | AMPAR backward rate constant | 190 / s | Destexhe et al. 1998 |

2.9 Post-synaptic $Ca^{2+}$ dynamics

The primary source of $Ca^{2+}$ in spines is NMDARs or VGCCs (Keller et al., 2008). In our model, there is no NMDAR present, hence the primary source of the $Ca^{2+}$ is the VGCC. Pharmacological analysis suggests that CA1 spines contain mostly R-type $Ca^{2+}$ channels (Sabatini et al., 2001; Bloodgood & Sabatini, 2007). Thus it is the only type of VGCC present on the surface of a spine-head in our model. AMPAR with missing GluR2-subunit are also permeable to $Ca^{2+}$. It has been reported that 1.2 – 1.3 % of the current carried by AMPAR is $Ca^{2+}$ current (Scheggenburger et al., 1993; Bollman et al., 1998). It is supposed that AMPAR is the only other source of $Ca^{2+}$ apart from VGCC. Apart from the above mentioned $Ca^{2+}$ sources, the effect of endogenous $Ca^{2+}$ buffers has also been incorporated in our model. These buffers are known to have a low buffering capacity of 20 units, which means ~5% $Ca^{2+}$ entering remains unbound (Helmchen, 2002). The buffering capacity of the endogenous buffers in our model is also 20 units. In fact, the endogenous buffer in our model is calmodulin, which binds with the cytosolic $Ca^{2+}$ to regulate the phosphorylation and the autophosphorylation of $Ca^{2+}$/calmodulin-dependent protein kinase (CaM-KII) which has been modeled in the next section.

$Ca^{2+}$ extrusion is supposed to be carried out with the help of the PMCa pump. The loss of the $Ca^{2+}$ due to the diffusion has not been considered (because of negligible diameter of neck). The reaction between $Ca^{2+}$ and endogenous buffer is modeled using the following bi-molecular reaction

$$c_{post} + b \underset{k_b}{\overset{k_f}{\rightleftharpoons}} c_{post} \cdot b. \tag{20}$$

Here $k_f$ and $k_b$ are the forward and backward rate constants respectively, $c_{post}$ represents the post-synaptic spine $Ca^{2+}$ concentration, $b$ represents the free buffer concentration and $c_{post} \cdot b$ represents bound $Ca^{2+}$ concentration. There are no sources or sinks for endogenous buffers. The reactions in equation (20) can be represented in the form of the following system of ordinary differential equations

$$\frac{dc_{post}}{dt} = f(c_{post}) - k_f c_{post}(b_t - c_{post} \cdot b) + k_b c_{post} \cdot b \tag{21}$$

$$\frac{dc_{post} \cdot b}{dt} = k_f c_{post}(b_t - c_{post} \cdot b) - k_b c_{post} \cdot b, \tag{22}$$

Here $f(c_{post})$ is the $Ca^{2+}$ due to the membrane influx and efflux and $b_t$ is the total endogenous buffer concentration. If we assume buffering kinetics to be fast, then we can take $c_{post} \cdot b$ to be in quasi-steady state i.e., $k_f c_{post}(b_t - c_{post} \cdot b) - k_b c_{post} \cdot b = 0,$ and solving for $c_{post} \cdot b$ we get (Higgins et al., 2006)

$$c_{post} \cdot b = \frac{b_t c_{post}}{K_{endo} + c_{post}}, \quad K_{endo} = k_b / k_f,$$

Then the total $Ca^{2+}$ concentration inside spine will be

$$\frac{dc_{post}}{dt} + \frac{dc_{post} \cdot b}{dt} = (1+\theta)\frac{dc_{post}}{dt} = f(c_{post}) \tag{23}$$

Where, $\theta = \dfrac{b_t K_{endo}}{(K_{endo} + c_{post})^2}$. The value of $K_{endo}$ has been taken to be that of calmodulin and $b_t$ was chosen so that the buffering factor $(1 + \theta)$ becomes 20 units i.e., equal to the buffering factor of endogenous buffers (Helmchen, 2002). Buffering effect, over post-synaptic $Ca^{2+}$ concentration, is approximated by division by the buffering factor

$$\frac{dc_{post}}{dt} = \frac{f(c_{post})}{1+\theta}. \tag{24}$$

As we know $f(c_{post})$ includes membrane proteins, VGCC and PMCa responsible for the $Ca^{2+}$ influx and efflux. The $Ca^{2+}$ current through VGCC is linear for the range of voltages in spines (-65 mV to -3 mV; Keller et al., 2008). We have assumed 12 VGCCs at the surface of spine, which is consistent with experimental estimates at the spine (1 – 20 VGCCs; Sabatini & Svoboda, 2000). Sabatini & Svoboda (2000) determined the mean open probability $P_{open}$ of the VGCC following a dendritic AP. Also in a spine containing $N$ channels, which opens with the probability $P_{open}$ per action potential, the number of channels opened by an action potential is governed by the binomial distribution (Katz & Miledi, 1970). Thus to determine the number of VGCCs open, we generated a binomially-distributed random number (BRN) whenever $V_{post}$ is greater than activation threshold of R-type channel i.e. – 30 mV (Berridge, 2009). The $Ca^{2+}$ current through the ensemble of R-type channels is

$$i_R = g_R B(N, P_{open})(V_{post} - V_R). \tag{25}$$

Here $g_R$ is the conductance of a single R-type channel, $B(N, P_{open})$ is the binomially distributed random variable determining the number of R-type channels open following post-synaptic neuron depolarization to -30 mV, $V_R$ is the reversal potential of hippocampal neuron R-type channel current determined by Sochivko et al (2002). The efflux through PMCa is taken to be of the following form

$$s_{pump} = k_s (c_{post} - c_{post}^{rest}). \tag{26}$$

Here $c_{post}^{rest}$ is the resting post-synaptic $Ca^{2+}$ concentration, $k_s$ is the maximum rate of $Ca^{2+}$ ion efflux due to the PMCa. Thus

$$f(c_{post}) = -\frac{(\eta I_{AMPA} + i_R)}{z_{Ca} F V_{spine}} - s_{pump}, \tag{27}$$

where $\eta$ is the fractional $Ca^{2+}$ current carried by AMPAR, $V_{spine}$ is the volume of the spine-head, $I_{AMPA}$, $i_R$, $s_{pump}$ are given by equation (19), equations (25)–(26). All other parameters are as stated in Table 8.

Table 8: List of parameters used for post-synaptic $Ca^{2+}$ dynamics

| Symbol | Description | Value | Reference |
|---|---|---|---|
| $\eta$ | Fraction $Ca^{2+}$ current carried by | 0.012 | Bollman et al., 1998 |

|  | AMPAR |  |  |
|---|---|---|---|
| $P_{open}$ | VGCC open probability | 0.52 | Sabatini & Svoboda, 2000 |
| $K_{endo}$ | $Ca^{2+}$ affinity of Endogenous buffer | 10 μM | Keller et al, 2008 |
| $b_t$ | Total endogenous buffer concentration | 200 μM | See text, Page No. 19 |
| $V_R$ | Reversal potential of $Ca^{2+}$ ion in spine | 27.4 mV | Sochivko et al, 2002 |
| $V_{spine}$ | Volume of dendrite spine | 0.9048 μm$^3$ | See text, Page No. 17 |
| $c_{post}^{rest}$ | Resting post-synaptic $Ca^{2+}$ concentration | 100 nM |  |
| $k_s$ | Maximum PMCa efflux rate | 100 / s | Keller et al. 2008 |
| $g_R$ | Conductance of R-type channel | 15 pS | Sabatini & Svoboda, 2000 |
| $N$ | Number of R-type channels | 12 | Sabatini & Svoboda, 2000 |

## 2.10 Post-synaptic CaM-KII phosphorylation

CaM-KII is a $Ca^{2+}$-activated protein kinase with switch-like properties (Griffith, 2004). CaM-KII consists of 8–12 subunits that form a petal structure (Zhabotinsky, 2000). The analysis of CaM-KII autophosphorylation and dephosphorylation indicate that it can serve as a molecular switch which is capable of long-term memory storage (Lisman et al., 2002). The role of CaM-KII autophosphorylation in induction of LTP is well accepted while its role in LTP maintenance is still an open question (Zhabotinsky, 2000). Phosphorylation of AMPAR by activated CaM-KII at Serine–831 (or Ser$^{831}$ is a residue of the GluR1 subunit of AMPAR) is the most observed form of LTP (Lee et al., 2003; Boehm & Malinow, 2005). The LTP we are trying to model does not involve the role of CaM-KII in phosphorylation of the AMPAR, as Perea & Araque (2007) observed no change in synaptic potency. The locus of induction of LTP observed by Perea & Araque (2007) was purely pre-synaptic requiring mild-depolarization of the post-synaptic neuron signifying the possible role of a retrograde messenger. NO might be the elusive retrograde messenger involved because of its demonstrated capability in regulating LTP at hippocampus (Bon & Garthwaite, 2003; Hopper & Garthwaite, 2006). Neuronal nitric oxide synthase (nNOS) enzyme is responsible for the production of NO inside the CNS (Calabrese et al., 2007). The production of NO from nNOS is known to be $Ca^{2+}$ and calmodulin dependent (Garthwaite & Boulton, 1995). Recently it has been demonstrated that CaM-KII phosphorylation leads to NO production from NOS in endothelial cells (Ching et al 2011). Also, both nNOS and CaM-KII coexist in the post-synaptic density (or PSD is a neuronal scaffolding protein that associates with post-synaptic membrane receptors and links them to intracellular signaling molecules, like kinases and phosphatases) (Nikonenko et al 2008) therefore a possible role of CaM-KII in NO production from nNOS cannot be overlooked. It is interesting to note that phosphorylation of nNOS at residue Serine-847 (Ser$^{847}$) by CaM-KII can decrease NO production in neuron cells (Watanbe et al., 2003) but phosphorylation of nNOS at residue Serine-1412 (Ser$^{1412}$) increases

NO production (Rameau et al., 2007). Rameau et al. (2007) demonstrated, with the help of *okadaic acid* a protein phosphatase 1 (PP1) inhibitor, that the phosphorylated CaM-KII may reduce dephosphorylation of nNOS at Ser$^{1412}$ and hence contribute in the NO production.

Based on the experimental observations discussed in the above paragraph, we infer that the Ca$^{2+}$ entering inside the post-synaptic spine-head through VGCC and AMPAR (as described in section 2.9) leads to phosphorylation of CaM-KII. This phosphorylated CaM-KII stimulates the production of NO from nNOS (McCue et al, 2010; Forstermann & Sessa, 2011). The produced NO is free to diffuse to the pre-synaptic terminal where it causes an increase in pre-synaptic PIP$_2$ concentration (Micheva et al, 2002). This PIP$_2$ prebinds with synaptotagmin (Ca$^{2+}$ sensor protein of the synaptic vesicles release machinery) and increases the affinity of the synaptotagmin for the boutonic Ca$^{2+}$ (Bai et al, 2004). This implies that increasing PIP$_2$ concentration enhances the rate with which Ca$^{2+}$ binds to the pre-synaptic Ca$^{2+}$ - sensor. The rate with which Ca$^{2+}$ binds to pre-synaptic Ca$^{2+}$ - sensor is given by Ca$^{2+}$ association rate constant ($\alpha$) (see equation (6)). $\alpha$ has been modeled to depend directly on phosphorylated CaM-KII concentration (due to the lack of exact mechanism by which CaM-KII affects PIP$_2$ via NO, and consequently the rate with which Ca$^{2+}$ binds with pre-synaptic Ca$^{2+}$ - sensor ).

We have used the mathematical model of CaM-KII phosphorylation due to Zhabotinsky (2000). The model consists of 10 subunits. When two neighboring subunits bind with (Ca$^{2+}$)$_3$CaM (where the subscript denotes 3 Ca$^{2+}$ ions), the first one phosphorylates the second one in clockwise direction. This initial step can be summarized by the following reactions

$$3Ca^{2+} + CaM \rightleftharpoons C, \qquad (28)$$

$$P_0 + C \rightleftharpoons P_0C, \qquad (29)$$

$$P_0C + C \rightleftharpoons P_0C_2, \qquad (30)$$

$$P_0C_2 \xrightarrow{K_1} P_1C_2. \qquad (31)$$

Here $C$ denotes Ca$^{2+}$ bound with calmodulin, $P_0$ is the unphosphorylated CaM-KII holoenzyme, $P_1$ is the one-fold phosphorylated CaM-KII holoenzyme. The rate of the initiation step or the rate of the phosphorylation is

$$v_{phos} = 10 K_1 P_0 \frac{\left(c_{post}^{n_{h_1}}\right)^2}{\left(k_{h_1}^{n_{h_1}} + c_{post}^{n_{h_1}}\right)^2}, \qquad (32)$$

where $K_1$ is the rate at which $P_0C_2$ is phosphorylated (see equation (31)), 10 is the statistical factor, $k_{h_1}$ is the post-synaptic Ca$^{2+}$ concentration at which $v_{phos}$ is halved and $n_{h_1}$ is the Hill coefficient for the rate of phosphorylation. Further, Zhabotinsky (2000) assumed that catalytic activity of the autophosphorylated subunit does not depend on the binding of $C$ and in this case the rate of autophosphorylation can be given by the following reactions

$$P_1 + C \rightleftharpoons P_1C, \tag{33}$$

$$P_1C \xrightarrow{K_1} P_2. \tag{34}$$

Correspondingly the per-subunit rate of autophosphorylation is

$$v_{\text{a-phos}} = K_1 \frac{c_{\text{post}}^{n_{h_1}}}{k_{h_1}^{n_{h_1}} + c_{\text{post}}^{n_{h_1}}}. \tag{35}$$

Here, $K_1$ is the rate constant of equation (34) which is the same as of equation (31). Dephosphorylation of subunits proceeds according to the following MM scheme

$$SP + E \rightleftharpoons SPE \longrightarrow E + S. \tag{36}$$

Here $S$ is an unphosphorylated subunit, $SP$ is a phosphorylated subunit, $E$ is a protein phosphatase and $SPE$ is the phosphorylated subunit bound with protein phosphatase. Following Zhabotinsky (2000) the per-subunit rate of dephosphoryaltion is

$$v_{\text{d-phos}} = \frac{K_2 e_p}{K_M + \sum_{1}^{10} i P_i}, \tag{37}$$

where $e_p$ is the concentration of active protein phosphatase, and $K_2$ and $K_M$ are catalytic and MM constants, respectively. Four protein phosphotases dephosphorylate CaM-KII: PP1, PP2A, PP2C and a specific CaM-KII phosphatase (Zhabotinsky, 2000). PP1 is the only protein which dephosphorylates CaM-KII in PSD (Zhabotinsky, 2000). Activity of PP1 can be controlled by $Ca^{2+}$/CaM via inhibitor 1 (I1), calcineurin (CaN) and protein kinase A (PKA). PKA phosphorylates I1 and CaN dephophorylates it (see Figure 2). Although, phosphorylated form of the inhibitor-1 (I1P) is known as a potent inhibitor of PP1 (by forming a complex PP1–I1P, which inactives the PP1 activity) (Zhabotinsky, 2000); its de-phosphorylated form is known to stimulate the PP1 activity (Munton, 2004).

Hence a modification was made to the equation governing PP1 concentration in Zhabotinsky (2000) to incorporate the I1 dependent PP1 production (see broken line Figure 2). Also recent values of PKA based phosphorylation of I1 were used (Sahin et al., 2006). The mathematical model governing CaM-KII phosphorylation can be given by the following equations (Lisman & Zhabotinsky, 2001)

$$\frac{dP_0}{dt} = -v_{\text{phos}} + v_{\text{d-phos}} P_1$$

$$\frac{dP_1}{dt} = v_{\text{phos}} - v_{\text{d-phos}} P_1 - v_{\text{a-phos}} P_1 + 2v_{\text{d-phos}} P_2$$

$$\frac{dP_i}{dt} = v_{\text{a-phos}} w_{i-1} P_{i-1} - v_{\text{d-phos}} i P_i - v_{\text{a-phos}} w_i P_i + v_{\text{d-phos}}(i+1) P_{i+1}$$

$$\frac{dP_{10}}{dt} = v_{\text{a-phos}} P_9 - v_{\text{d-phos}} 10 P_{10}$$

$$\frac{de_p}{dt} = -k_F I e_p + k_B (e_{p0} - e_p) + k_I I_0$$

$$\frac{dI}{dt} = -k_F I e_p + k_B (e_{p0} - e_p) + v_{\text{PKA}} \frac{I_0}{I_0 + K_{\text{PKA}}} - v_{\text{CaN}} I \frac{c_{\text{post}}^3}{k_{h_2}^3 + c_{\text{post}}^3}$$

(38)

Here $P_i$ is the i-fold phosphorylated CaM-KII holoenzyme, $e_p$ is the concentration of PP1 not bound to I1P, $e_{p0}$ is the total concentration of PP1, $I$ is the concentration of free I1P and $I_0$ is the concentration of free I1. If 'i' subunits are phosphorylated then the effective number of subunits capable of autophosphorylating are $w_2 = w_8 = 1.8$, $w_3 = w_7 = 2.3$, $w_4 = w_6 = 2.7$, $w_5 = 2.8$ (here subscript is 'i'). $k_F$ and $k_B$ are respectively the association and dissociation rate constant of PP1–I1P complex. $k_{h_2}$ is the Hill constant of CaN dependent dephosphorylation of I1P, $v_{\text{CaN}}$ is the rate of dephosphorylation of I1P due to CaN, $K_{\text{PKA}}$ is the MM constant of PKA, $v_{\text{PKA}}$ is the rate of phosphorylation of I1 by PKA, $k_I$ is an experimentally undetermined rate, with which I1 regulates PP1 activity, fixed with the help of computer simulations. The importance of incorporating I1 dependent PP1 production is that the CaM-KII phosphorylation process becomes reversible, rather than bi-stable (having two stable steady-states), as found experimentally by Bradshaw et al (2003). The value of the parameters is listed in Table 9. For more details on the model see Zhabotinsky (2000).

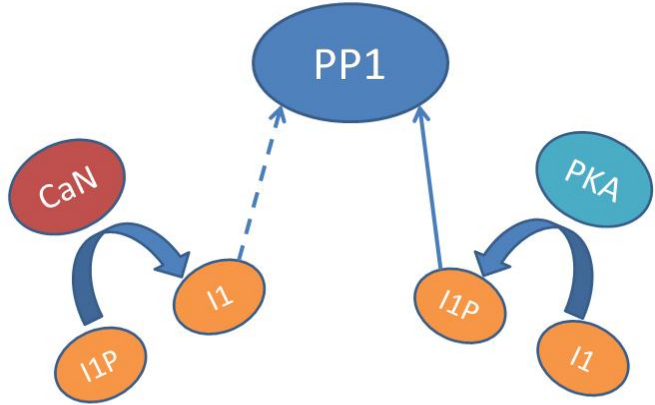

Figure 2. Schematic representation of PP1 activity regulation by CaN and PKA. CaN dephosphorylates I1P and promotes PP1 activity via I1 (broken line). PKA phosphorylates I1 and inhibits PP1 activity via I1P (solid line).

Table 9: List of parameters used for CaM-KII Phosphorylation

| Symbol | Description | Value | Reference |
|---|---|---|---|
| $k_F$ | Association rate constant of PP1 – I1P complex | 1 / μM s | Zhabotinsky, 2000 |
| $k_B$ | Dissociation rate constant of PP1 – I1P complex | $10^{-3}$ / s | Zhabotinsky, 2000 |
| $k_I$ | I1 dependent regulation rate of PP1 | 1 / s | See text, Page No. 24 |
| $e_{p0}$ | Total concentration of PP1 | 0.1 μM | Zhabotinsky, 2000 |
| $v_{PKA}$ | Phosphorylation rate of I1 due to PKA | 0.45 μM / s | Sahin et al., 2006 |
| $K_{PKA}$ | MM constant of I1 phosphorylation rate | 0.0059 μM | Sahin et al., 2006 |
| $v_{CaN}$ | Rate of I1P dephosphorylation due to CaN | 2 / s | Lisman & Zhabotinsky, 2001 |
| $k_{h_2}$ | $Ca^{2+}$ activation Hill constant of CaN | 0.7 μM | Zhabotinsky, 2000 |
| $K_1$ | Catalytic constant of autophosphorylation | 0.5 / s | Zhabotinsky, 2000 |
| $k_{h_1}$ | $Ca^{2+}$ activation Hill constant of CaM-KII | 4.0 μM | Zhabotinsky, 2000 |
| $n_{h_1}$ | Hill coefficient for $Ca^{2+}$ activation of CaM-KII | 3 | Miller et al, 2005 |
| $K_2$ | Catalytic constant of PP1 | 10 / s | Bradshaw et al, 2003; Miller et al, 2005 |
| $K_M$ | MM constant of PP1 | 20 μM | Zhabotinsky, 2000 |
| $I_0$ | Concentration of free I1 | 0.1 μM | Zhabotinsky, 2000 |
| $k_{syt}$ | Percent increase in $Ca^{2+}$ association rate of synaptotagmin | 0.5% | See text; Page No. 25 |
| $P_{1/2}$ | Threshold for phosphorylated CaM-KII based NO production | 40 μM | See text; Page No. 26 |
| $k_{1/2}$ | Slope factor for activation of CaM-KII based NO production | 0.4 μM | See text; Page No. 26 |
| $e_k$ | Total CaM-KII concentration in PSD | 80 μM | Zhabotinsky, 2000 |

As explained earlier, it is assumed that the phosphorylated CaM-KII stimulates NO production. Produced NO is free to diffuse to the pre-synaptic terminal, where it enhances the rate with which $Ca^{2+}$ binds with the synaptic vesicle $Ca^{2+}$ - sensor. Mathematically this has been achieved by increasing the rate with which $Ca^{2+}$ binds with $Ca^{2+}$ - sensor (synaptotagmin) i.e. increasing the value of $Ca^{2+}$ association rate (α) with $Ca^{2+}$ - sensor. The percent increase in the $Ca^{2+}$ association rate (α) is an important parameter and has been termed as $k_{syt}$. Only a small increment of 5% in α value could yield us a persistent increase, up to 80%, in synaptic efficacy. The increase in α value is governed by a sigmoidal function given by

$$\frac{k_{syt}}{1+\exp\left(-\frac{\left(\left(\sum_{1}^{10} P_i\right) - P_{1/2}\right)}{k_{1/2}}\right)}. \tag{39}$$

Here $P_{1/2}$ is the (choosen to be half of the total CaM-KII concentration inside PSD) threshold value of phosphorylated CaM-KII after which it stimulates NO production, $k_{1/2}$ is the slope giving the increment a switch like behavior which is dependent upon the phosphorylated CaM-KII concentration. It is apparent from equation (39) that the increment can achieve a maximum of $k_{syt}$ for a fully phosphorylated CaM-KII. We simulated the model for varying values of $k_{syt}$ and demonstrated the change in LTP for pre-astrocytic to post-astrocytic activities. The value of the parameters is appended in Table 9.

2.11 Numerical Implementation

All the computations and visualizations of the model are implemented in MATLAB environment. The model equations were discretized with a temporal precision of, Δt = 0.05 ms. The canonical explicit Euler method was used to solve the system of thirty-six ordinary differential equations governing TpS. For the numerical simulation of the noise term, in equation (12), we have used Box-Muller Algorithm (Fox, 1997) to generate the noise-term at each time-step (Δt). All simulations were performed on a Dell Precision T3500 workstation with Intel Xeon processor (4 parallel CPUs) with 2.8 GHz processing speed and with 12 GB RAM. Simulating real time of 1 minute takes approximately 9.5 minutes of simulation time. The MATLAB script written for the simulation of the model is supplied with the Supporting Material.

## 3. Simulation Results

3.1 Astrocyte-independent information processing

We first demonstrate how the model would behave in an astrocyte-independent pathway (see Figure 1). The results are shown for a typical simulation elicited in response to a stimulus current of density 10 μA cm$^{-2}$ (frequency 5 Hz, duration 10 ms), imitating the single pulses stimulus used by Perea & Araque (2007) in their experiments. The experiments performed by Perea & Araque (2007) were considering an open loop i.e. simultaneous stimulation of SC and astrocyte. Such parallel stimulation of SC and astrocyte can occur at a TpS under physiological condition. But the pharamacological stimulus of astrocyte used in their experiments may or may not e.g., UV-flash stimulated NP-EGTA mediated Ca$^{2+}$ uncaging in astrocytes at first appears to be an excellent approach to study Ca$^{2+}$ - dependent processes. However there are caveats associated with this approach (Agulhon et al, 2008). Thus in our model we have considered a closed loop, where astrocyte is excited in an activity-dependent manner i.e. astrocytic Ca$^{2+}$ rises in response to the glutamate released in the cleft from SC.

Asking the question of the effect of an astrocyte over the synaptic plasticity, Perea & Araque (2007) first only stimulated the SC and not the astrocyte. The present section is analogous to their experimental setup studying synaptic efficacy, synaptic potency and Pr in absence of astrocyte. In Figure 2, important components like pre-synaptic membrane potential, bouton $Ca^{2+}$ concentration, synaptic glutamate and post-synaptic membrane potential are shown. The model was simulated for a duration of 30 minutes. But the results in Figure 3 have been shown for 60 seconds just for the sake of clarity of the involved variable.

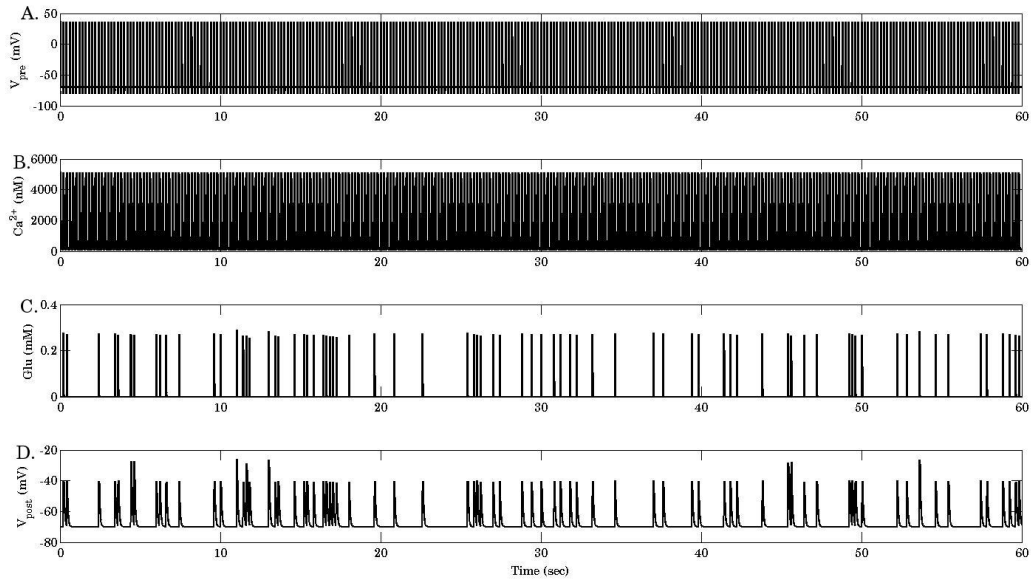

Figure 3. Components involved in astrocyte-independent pathway of information processing. A. Pre-synaptic AP, generated at axon-hillock in response to the stimulus. B. $Ca^{2+}$, as a result of AP invading axon boutons. C. Glutamate, released in the synaptic cleft due to exocytosis of vesicles in a $Ca^{2+}$ - dependent manner. D. Excitatory Post-Synaptic Potential (EPSP), due to opening of ionotropic glutamate receptors (here, AMPARs) on the surface of spine-head.

To simulate the astrocyte-independent information flow from the pre-synaptic bouton to the post-synaptic spine, we clamped astrocyte $Ca^{2+}$ to its resting value regardless of the stimulus (in the form of released synaptic glutamate) received at astrocytic mGluR. Figure 3A is the pre-synaptic AP generated at the axon-hillock in response to the current injected at soma. As assumed, AP propagates without degradation to the axon bouton leading to the opening of the VGCCs. The $Ca^{2+}$ current carried by the VGCC leads to an increase in the bouton $Ca^{2+}$ concentration (see Figure 3B). The $Ca^{2+}$ inside the bouton is generally released in close proximity of the synaptic vesicles which binds with its $Ca^{2+}$ - sensor instigating the process of exocytosis (see Figure 3C). The released glutamate is free to diffuse and binds with glutamate receptors (AMPAR on the post-synaptic spine and mGluR on the astrocyte surface). Since astrocyte $Ca^{2+}$ was clamped to its resting concentration there is no amplification of the information (see Figure 1) due to it and hence omitted. Glutamate bound with AMPARs causes depolarization of post-synaptic membrane (see Figure 3D). Figure 3D basically reflects the transduction of the input signal (see Figure 3A). It should be noted that the given process of signal transduction does not make any contribution due to the astrocyte-dependent pathway. The synaptic strength for the present synapse is calculated in a way analogous to Perea & Araque (2007). We calculated synaptic efficacy (windowed-mean of post-synaptic responses including success and failures), synaptic potency (windowed-mean of successful post-synaptic responses) and Pr (given as a measure of signal transduction i.e. ratio of the input pulse being converted into the output response). Data are expressed as mean ± standard deviation. The results were compared using a two-tailed Student's t-test. Statistical significance is established with $P < 0.01$.

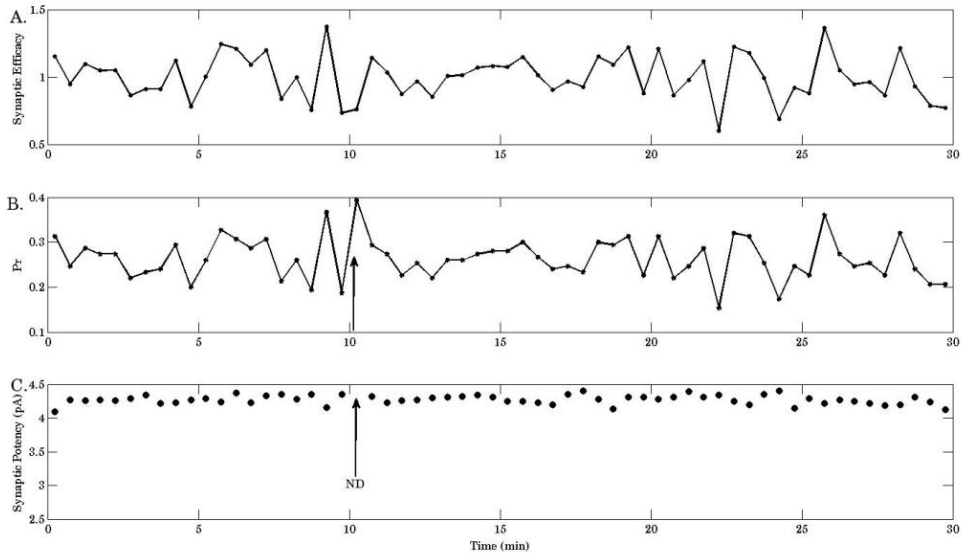

Figure 4. Synaptic parameters given as synaptic efficacy, probability of neurotransmitter release and synaptic potency. The black arrow (↑) is to show the time at which post-synaptic neuron was depolarized to -30 mV; ND: Neuronal Depolarization. A. Synaptic efficacy, B. Probability of neurotransmitter release and C. Synaptic potency has been averaged over a time–window of 30 seconds.

It is apparent from figure 4 that the synaptic efficacy remains unchanged prior to post– synaptic neuron depolarization (to − 30 mV) and after 10 minutes of post-synaptic depolarization (before: 0.029 ± 0.005 pA; after 10 minutes: 0.028 ± 0.005 pA; $P = 0.51$). Similarly, probability of neurotransmitter release Pr remained unchanged prior to the post-synaptic depolarization and 10 minutes after post–synaptic depolarization (before: 0.26 ± 0.05; after 10 minutes: 0.25 ± 0.05; $P = 0.53$). Synaptic potency also remained unperturbed (before: 4.27 ± 0.07 pA; after 10 minutes: 4.26 ± 0.07 pA; $P = 0.71$) prior to post-synaptic depolarization and 10 minutes after post-synaptic depolariztion. Our observations are in agreement with the experimental observations of Perea & Araque (2007), when peri-syanptic astrocyte was not photo-simulated, which validates our model at an astrocyte-independent synapse.

In Figure 5A, we show post-synaptic spine $Ca^{2+}$ concentration due to the fractional current carried by AMPARs and VGCCs. Also, one can see a cluster of $Ca^{2+}$ peaks at the 10 minutes mark, which is in response to post-synaptic membrane depolarization to $-30$ mV. Whenever post-synaptic membrane is depolarized (for more than $-30$ mV) they lead to opening of VGCCs (depending upon $B(N, P_{open})$, see equation (25)) located at the spine. VGCCs are responsible for the cluster of $Ca^{2+}$ peaks visible in Figure 5A. Inflow of excess $Ca^{2+}$ through VGCCs is good enough to phosphorylate CaM-KII above the threshold ($P_{1/2}$), see Figure 5B. The transient increase in Pr, visible in Figure 4B, is as a result of CaM-KII crossing the threshold (see Figure 5B). However, in absence of astrocyte the frequency of the post-synaptic $Ca^{2+}$ response is not strong enough to keep it phosphorylated above $P_{1/2}$. Consequently, CaM-KII is de-phosphorylated below the threshold and Pr comes down to values prior to the post-synaptic depolarization.

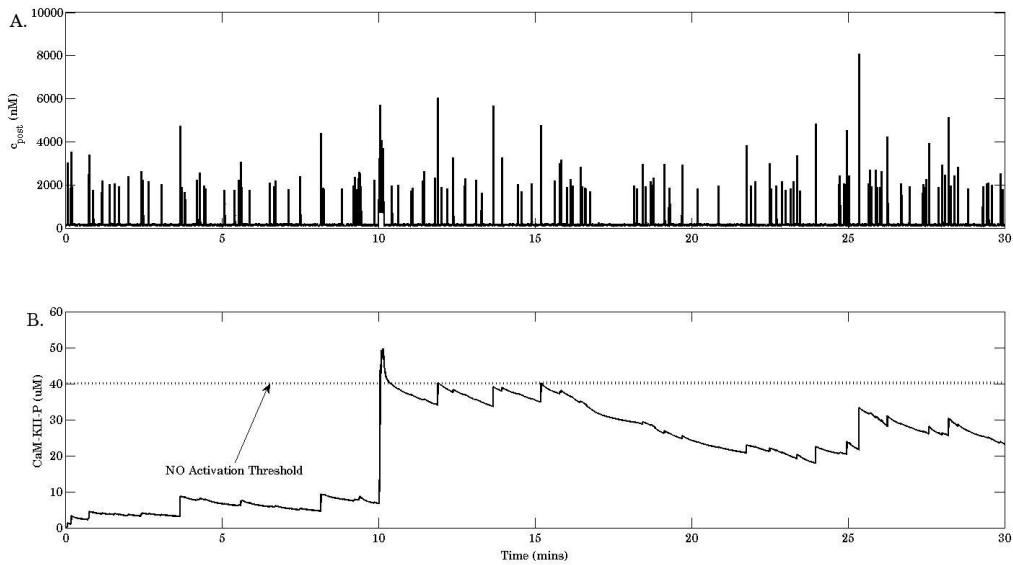

Figure 5. Post-synaptic variables responsible for stimulating NO production in our model. (A) Post-synaptic spine $Ca^{2+}$ concentration. (B) Phosphorylated CaM-KII (CaM-KII-P) concentration. Note the sudden spike in phosphorylated CaM-KII concentration at 10 minutes which is in response to mild-depolarization of post-synaptic neuron to $-30$ mV for duration of 10 seconds.

Figure 6 displays the evolution of the variables involved in CaM-KII phosphorylation process see equation (38). First we show the evolution of 0-fold phosphorylated CaM-KII holoenzyme, $P_0$ in equation (38), which is equal to the total CaM-KII concentration ($e_k$) minus phosphorylated CaM-KII concentration (i.e. equal to the sum of all i-fold phosphorylated CaM-KII holoenzymes or $\sum_{i=1}^{10} P_i$) shown in Figure 5B. It is interesting to note that, here, CaM-KII dephosphorylation process follows the scheme shown in equation (36). The relationship between 0-fold phosphorylated CaM-KII (or unphosphorylated CaM-KII) and phosphorylated CaM-KII is straightforward i.e. $P_0 = e_k - \sum_{i=1}^{10} P_i$. It is solely because in PSD CaM-KII concentration is much more than PP1 concentration, because of which the amount of CaM-KII bound with PP1 is negligible compared to the total CaM-KII concentration. This condition is also a reason behind the choice of MM kinetics for CaM-KII dephosphorylation process by Zhabotinsky (2000). PP1 (see Figure 6C) is deactivated by phosphorylated I1 (see Figure 6B). When I1P concentration is the lowest then PP1 concentration is the highest (at the 10 minutes mark). In figure 6C we can see a sharp spike in PP1 concentration which is due to a rapid fall in I1P concentration contributed by CaN via the cluster of $Ca^{2+}$ peaks (see Figure 5A). Evolution of all i-fold phosphorylated CaM-KII holoenzymes ($P_i$) is provided with the supplementary material.

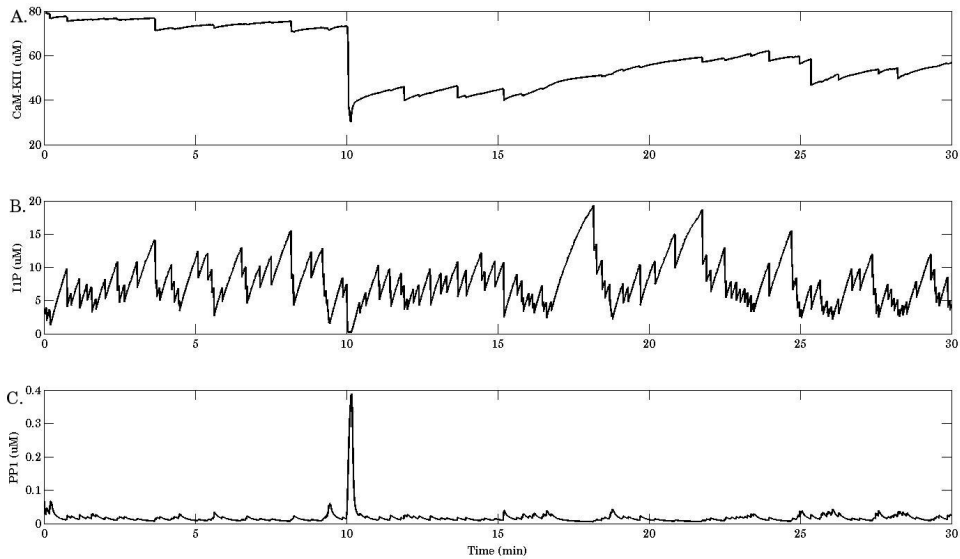

Figure 6. Some important variables involved in CaM-KII phosphorylation. A. 0-fold phosphorylated CaM-KII concentration, B. Phosphorylated I1 concentration, C. Free PP1 concentration (not bound to I1P).

3.2 Astrocyte-dependent information processing

We went ahead and investigated if clamping and unclamping of astrocytic $Ca^{2+}$ concentration would yield modifications similar to Perea & Araque (2007). They stimulated SC and astrocyte simultaneously to study the astrocytic role in synaptic strength modulation i.e. considering an open loop. However we considered a closed loop where astrocytic $Ca^{2+}$ is modulated by pre-synaptically released glutamate (Porter & McCarthy, 1996) which in turn modulates pre-synaptic $Ca^{2+}$ in a mGluR-dependent pathway (Fiacco & McCarthy, 2004; Perea & Araque, 2007). To validate the fact that astrocytes modulate synaptic strength via an mGluR-dependent pathway Perea & Araque (2007) continuously stimulated SC and evoked astrocytic $Ca^{2+}$ responses, using UV-flash via NP-EGTA, to find a transient enhancement of synaptic efficacy when astrocyte was photo-stimulated. Similar to their experimental approach we kept astrocytic $Ca^{2+}$ concentration clamped for the first 10 minutes imitating the behavior of NP-EGTA in their experiments. The astrocyte $Ca^{2+}$ clamping is turned off after 10 minutes. At the same time post-synaptic membrane is also depolarized to − 30 mV.

In Figure 7 we observe the changes in the model following an astrocyte-dependent approach of information processing. The release of $Ca^{2+}$ from bouton stores (ER) contributes to the transient increase in $[Ca^{2+}]$ (which is held responsible for modulating spontaneous transmitter release - Emptage et al. (2001)). With the information flow following an astrocyte-dependent approach, it is assumed that the agonist (glutamate, necessary for $Ca^{2+}$ release from ER), which causes cleavage (in a G-protein dependent manner) of $PIP_2$ into $IP_3$ and DAG, comes from the peri-synaptic astrocyte. Here we can observe that there is no change in bouton $IP_3$ concentration till it reaches 10 minutes mark (see Figure 7A). It is because of an assumed clamping of astrocytic $Ca^{2+}$ by NP–EGTA.

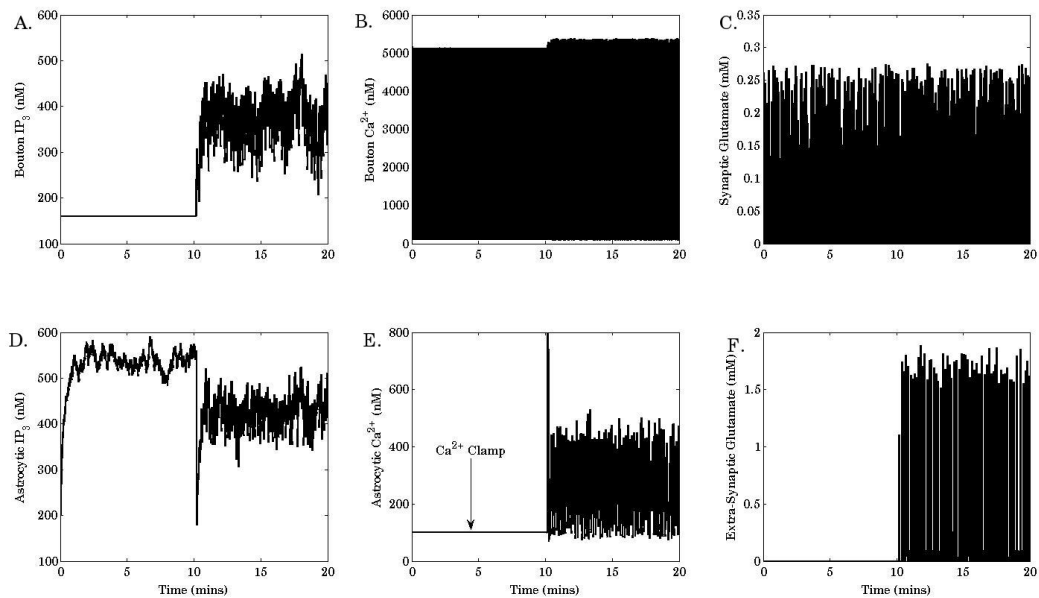

Figure 7. Components involved in astrocyte-dependent information processing. A. Bouton $IP_3$, B. Bouton $Ca^{2+}$ averaged in a time-window of length 30 seconds, C. Synaptic Glutamate, in response to exocytosis of synaptic vesicles, D. Astrocytic $IP_3$, due to synaptic glutamate binding with mGluRs on the surface of astrocyte, E. Astrocyte $Ca^{2+}$, elicited in response to $Ca^{2+}$ released from ER, F. Extra-synaptic glutamate, due to a possible exocytosis of SLMVs.

As soon as we remove the astrocytic $Ca^{2+}$ clamp, we can see an increase in $IP_3$ concentration, which opens the $IP_3R$ on the surface of ER. Further flow of $Ca^{2+}$ ions inside bouton cytosol due to release of $Ca^{2+}$ from ER elevates the averaged $Ca^{2+}$ concentration (see Figure 9A). $Ca^{2+}$ binds with $Ca^{2+}$ - sensor of the vesicles and instigates exocytosis of synaptic vesicles filled with glutamate (see Figure 7C). In a process similar to the one defined earlier, the synaptic glutamate leads to production of $IP_3$ but this time inside the astrocyte. Produced $IP_3$ binds with the $IP_3R$ which sets up inflow of $Ca^{2+}$ from the ER into the astrocytic cytosol (see Figure 7E). Astrocytic $Ca^{2+}$ binds with its $Ca^{2+}$ - sensor synaptotagmin IV protein (Montana et al., 2006) to initiate exocytosis of the SLMV's content (glutamate) in the extra-synaptic cleft (see Figure 7E). This spilled extra-synaptic glutamate is the necessary "agonist" mentioned in the earlier paragraph, which is responsible for the fluctuations of the bouton $IP_3$ (see Figure 7A) setting up enhanced bouton $Ca^{2+}$ concentration (see Figure 7B).

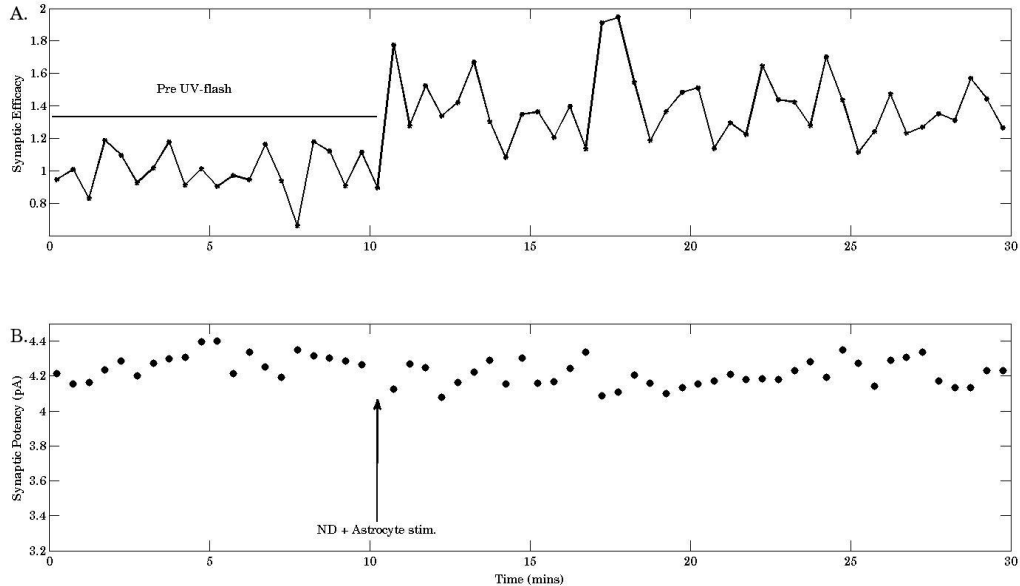

Figure 8. Synaptic parameters namely, synaptic efficacy and synaptic potency given as a measure of synaptic plasticity in an astrocyte-dependent approach of information processing. A. Synaptic efficacy is given by the average of EPSC amplitudes (including successes and failures) relative to the mean of EPSC amplitudes observed during the pre UV-flash period. Averages have been taken with a time-window of length 30 seconds, B. Synaptic potency is given by the average of the successful post-synaptic responses using a time-window of length 30 seconds. The arrow (↑) indicates the time when astrocyte was switched on ($Ca^{2+}$ unblock) and post-synaptic membrane was depolarized to – 30 mV. ND = neuron depolarization.

The effect of astrocyte-dependent pathway over the bouton $Ca^{2+}$ is pretty evident in Figure 7, but if this increase is playing any role at all in modulating the synaptic parameters, we that's what investigated next. We calculated synaptic efficacy, probability of neurotransmitter release and synaptic potency, as defined previously, using a time-window of length 30 seconds. We again made use of two-tailed student's t-test and established statistical significance (with $P < 0.01$) as defined earlier. The statistical inference did support the fact that astrocytes modulate synaptic parameters. The synaptic efficacy exhibited a persistent (≥20 minutes) LTP of up to 180% after mild post-synaptic neuron depolarization and astrocytic $Ca^{2+}$ unblock (from $0.028 \pm 0.003$ pA to $0.038 \pm 0.004$ pA ($P < 0.001$), before and 10 minutes after the astrocyte stimulation, respectively). In the experiments of Perea & Araque (2007) the locus of LTP induction was purely pre-synaptic, which was confirmed with the two tailed t-test of persistently unchanged synaptic potency. Similar to them we also performed a two tailed t-test on the averaged synaptic potency prior to the post-synaptic depolarization and 10 minutes after the post-synaptic depolarization. There was no statistically significant change in synaptic potency after LTP induction (4.25 ± 0.07 pA and 4.23 ± 0.07 pA, before and 10 minutes after the astrocytic stimulation; $P = 0.43$) validating the pre-synaptic locus of induction.

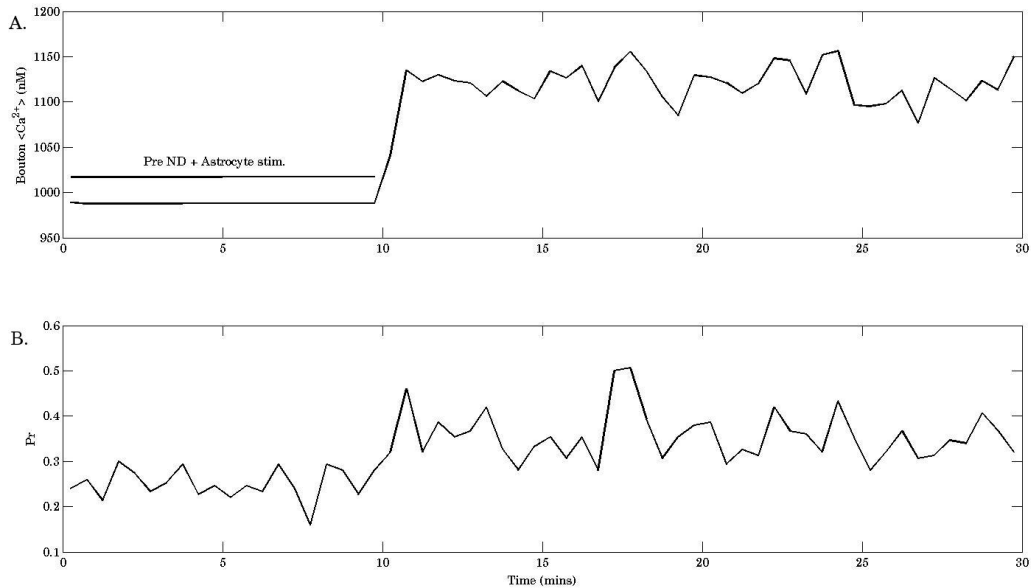

Figure 9. A. Averaged bouton $Ca^{2+}$ concentration. B. Probability of neurotransmitter release. The time-window for calculating the average was 30 seconds long in both the cases. Pre ND + astrocyte stim represents the phase prior to the post-synaptic membrane depolarization and astrocytic $Ca^{2+}$ unblock.

Next we checked if the persistent increase in probability of neurotransmitter release is associated with this form of LTP. Indeed we found a persistent increase in Pr (from 0.25 ± 0.03 to 0.35 ± 0.04; $P < 0.001$), before and 10 minutes after the astrocyte stimulation. This persistent increase in probability is due to an enhanced bouton average $Ca^{2+}$ concentration, visible after astrocyte $Ca^{2+}$ unblock (see Figure 9A). Another important reason related with this persistent probability increase is the mild post-synaptic depolarization to – 30 mV. The depolarization of post-synaptic membrane to – 30 mV causes VGCCs located at the spine to open, which brings additional $Ca^{2+}$ (apart from the fraction of $Ca^{2+}$ carried by AMPARs) inside the spine (see Figure 10A for a cluster of $Ca^{2+}$ peaks). This cluster of $Ca^{2+}$ peaks (at the 10 minutes mark), instigated by the post-synaptic depolarization, plays an important role in phosphorylating CaM-KII above the threshold (see Figure 10B) and stimulates NO production (as explained in subsection 2.10). We followed the similar procedure to stimulate NO production in astrocyte-independent pathway of information processing, but in vain, as frequency of post-synaptic spine $[Ca^{2+}]$ was unable to keep CaM-KII phosphorylated. However it is apparent from Figure 10B that in an astrocyte-dependent pathway a similar post-synaptic depolarization leads to a persistent increase in Pr and consequently, a persistent increase in synaptic efficacy, termed as LTP. It should be pointed out that in both the cases post-synaptic membrane was depolarized (to – 30 mV) for a duration of 10 seconds. The opening of VGCCs was governed by a BRN therefore a similar stimulus will open different number of VGCCs each time. Therefore for the sake of a fair comparison between the two pathways (astrocyte-dependent and astrocyte-independent) we ran simulations so that CaM-KII is phosphorylated (above the threshold for NO production) to the same extent for both pathways (as it will be evident from Figure 5B and 10B. It is ≈50 μM in both the cases). It should also be stated that the difference between the two-pathways remained apparent regardless of the extent to which CaM-KII is phosphorylated in response to a post-synaptic membrane depolarization. However it did modify the time with which CaM-KII decays below $P_{1/2}$ with stronger depolarization increasing time required for CaM-KII, to decay below $P_{1/2}$.

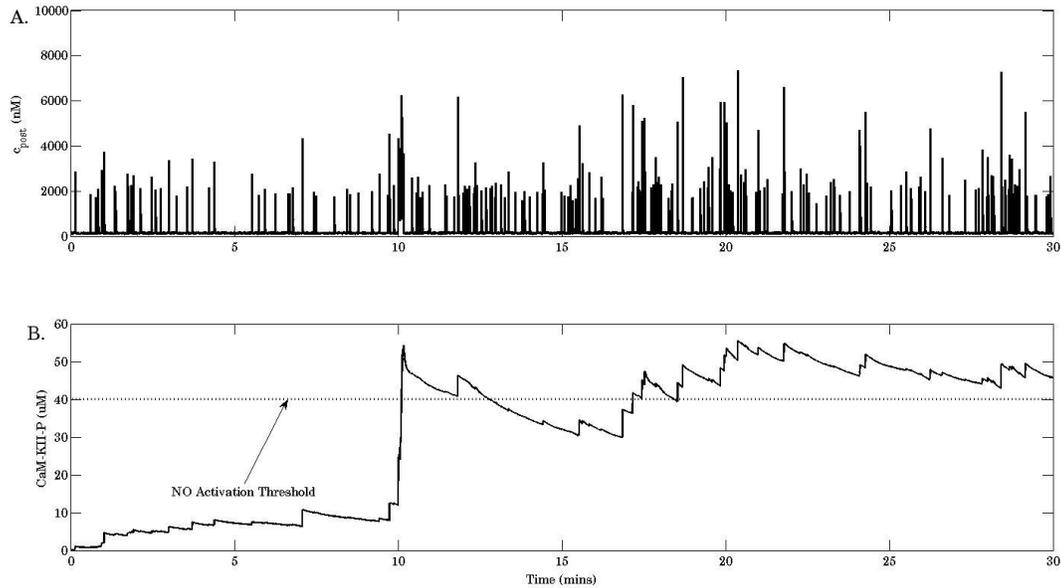

Figure 10. Post-synaptic variables responsible for stimulating NO production in our model. A. Post-synaptic spine $Ca^{2+}$ concentration, B. Phosphorylated CaM-KII (CaM-KII-P). The arrow (↑) points to the NO activation threshold, chosen to be the half of the total CaM-KII concentration inside PSD.

In any case, when CaM-KII decays below $P_{1/2}$ under astrocyte-independent pathway it stays below it (see Figure 5B). However even if it goes below $P_{1/2}$ following astrocyte-dependent pathway it again goes above $P_{1/2}$ (see Figure 10B), because of more frequent post-synaptic $Ca^{2+}$ responses due to enhanced average bouton $Ca^{2+}$ concentration (see Figure 9A). There was a clear distinction in CaM-KII phosphorylation process when astocyte was and wasn't photo-stimulated. Even the extent to which CaM-KII is phosphorylated was quite evident and $P_{1/2}$ could be fixed accordingly to predict the threshold for NO production (tested for $K_M$ (M–M constant of PP1 dependent dephosphorylation of CaM-KII) values of 2.5 μM, 5 μM, and 10 μM while keeping all other parameters fixed (see Figure 11)). It can be observed from Figure 11, that the threshold value of NO production ($P_{1/2}$) can always be fixed to be twice the MM constant of PP1 ($K_M$) dependent dephosphorylation of CaM-KII. The threshold of NO production basically defines the extent to which CaM-KII can be phosphorylated in an astrocyte-independent pathway of information processing (see the dotted line in Figure 11).

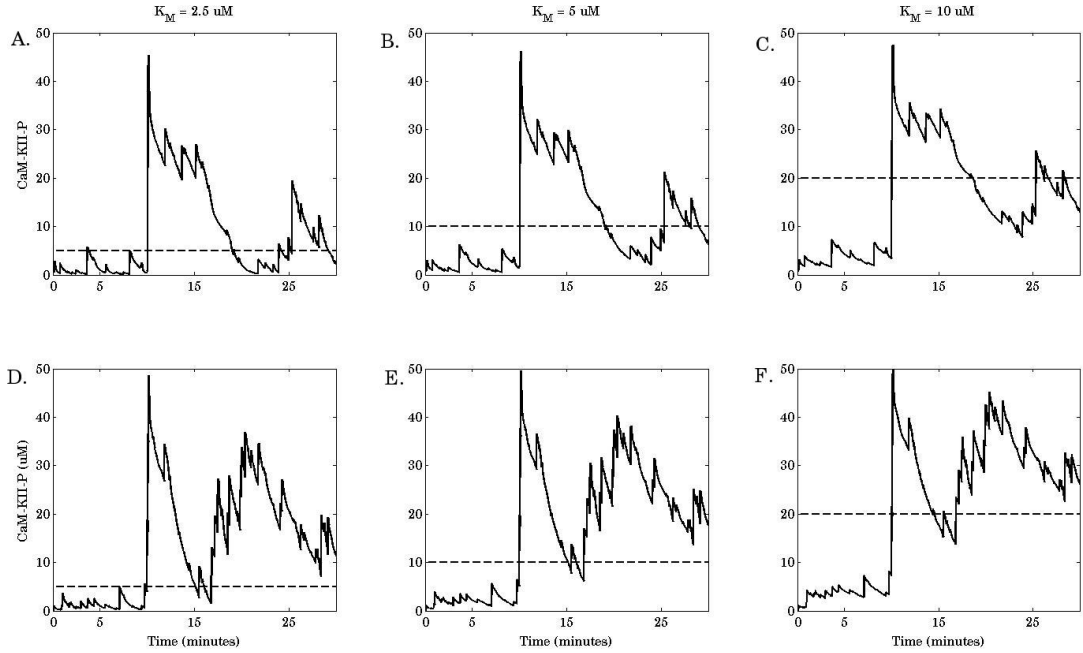

Figure 11. Evolution of CaM-KII phosphorylation process for different values of $K_M$. Dotted line represents the NO production threshold set to be 5 μM, 10 μM and 20 μM for $K_M$ values of 2.5 μM, 5 μM and 10 μM respectively. A – C are for an astrocyte-independent process. D – F are for an astrocyte-dependent process. (A, D), (B, E) and (C, F) are for $K_M$ = 2.5 μM, 5 μM and 10 μM, respectively.

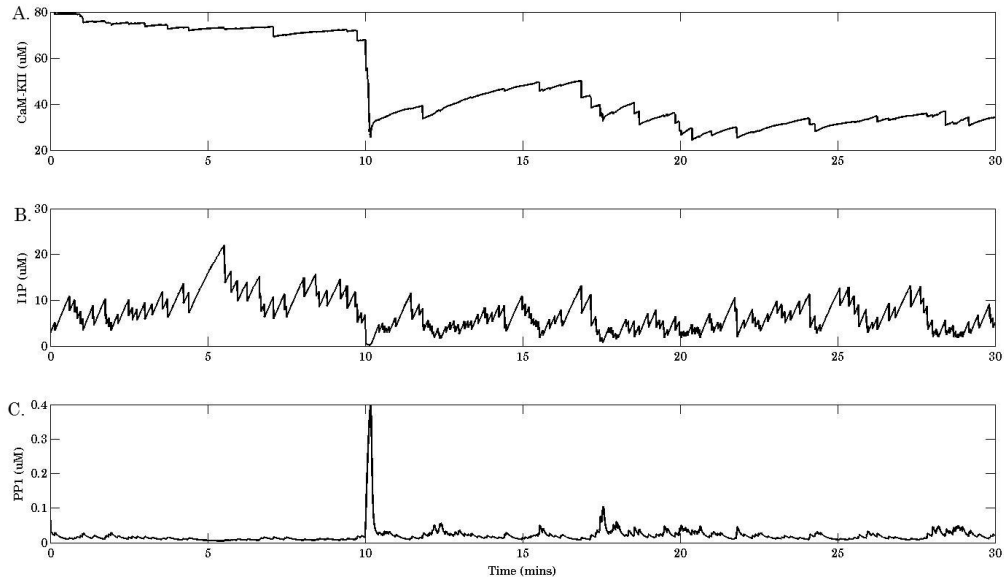

Figure 12. Some important variables involved in CaM-KII phosphorylation process. A. 0-fold phosphorylated CaM-KII, B. Phosphorylated I1 (I1P), C. Free PP1 i.e. not bound to I1P.

In Figure 12 we show some important variables associated with CaM-KII phosphorylation process performed mathematically using equation (38). As explained earlier the relationship between 0-fold phosphorylated CaM-KII and phosphorylated CaM-KII is straightforward. It can also be observed from Figure 12A that it is very similar to Figure 10B however inverted. PP1 (see Figure 12C) is deactivated by I1P (see Figure 12B). Following Zhabotinsky we assumed the concentration of free I1 as constant (at 0.1 µM). But one can observe from Figure 12B (and likewise from Figure 6B) that the concentration of phosphorylated I1 i.e. I1P is well above the free I1 concentration of 0.1 µM. Actually, the total concentration of I1 in PSD is significantly greater than free I1 (Miller et al, 2005) which is the case in our simulations. Evolution of all i-fold phosphorylated CaM-KII holoenzymes ($P_i$) is provided with the supplementary material.

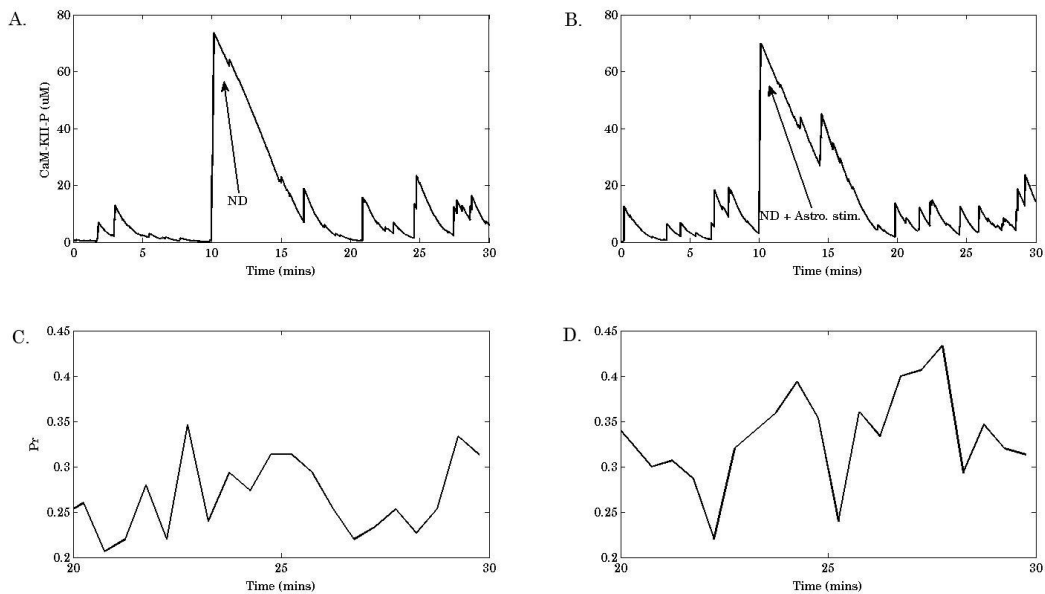

Figure 13. The effect of post-synaptic neuron depolarization (ND) and astrocyte stimulation in absence of I1 protein. A. Phosphorylated CaM-KII (CaM-KII-P) while assuming NP-EGTA block of astrocyte $Ca^{2+}$. B. Phosphorylated CaM-KII (CaM-KII-P) when NP-EGTA block is removed at 10 minutes mark. C. Neurotransmitter release probability (Pr) 10 minutes after ND, D. Neurotransmitter release probability (Pr) 10 minutes after ND and NP-EGTA unblock. Pr was calculated as previously using a time-window of 30 seconds. We have shown the figure 13C and 13D from 25 minutes to 30 minutes only for the sake of clarity.

The results discussed till now help establish our mathematical model, governing signal transduction, with and without the aid of a peri-synaptic astrocyte. A possible role of astrocyte in LTP induction has been discussed and a mathematical framework under which it occurs is presented. However there are other important signaling cascades that can abolish LTP. One such important protein is I1 which deactivates PP1 and helps in LTP induction. A mice lacking I1 protein has been shown to have deficits in LTP induction (Bibb et al, 2001). Figure 13 demonstrates how such a mutation can abolish LTP in our model.

Mathematically we have achieved this by setting $I_0 = 0$ in equation (38). This results in $e_p = e_{p_0}$ and is equivalent to removing the last two equations in the equation (38) and treating $e_p$ as a parameter. In Figure 13 the value of $e_p$ is equal to 0.05 μM. All other parameters are the same as listed in Tables 2–9. As expected removal of I1 from the PSD stops the I1 dependent deactivation of PP1 (which is responsible for dephosphorylating CaM-KII). Consequently PP1 continuously dephosphorylates CaM-KII (see Figure 13A-13B). In figure 13A we depolarized the post-synaptic membrane (to − 30 mV) and observed the change in neurotransmitter release probability. There was no statistically significant change in Pr before ND and 10 minutes after it (before: $0.28 \pm 0.03$; 10 minutes after: $0.26 \pm 0.04$; $P = 0.15$). We can observe from Figure 13B, that even simultaneous stimulus of ND and astrocyte $Ca^{2+}$ unblock could not keep CaM-KII-P above $P_{1/2}$ contrary to the behavior observed in figure 10B. However, we could observe a statistically significant change in Pr before stimulus (ND + astro. stim.) and 10 minutes after it (before: $0.26 \pm 0.04$; 10 minutes after: $0.33 \pm 0.05$; $P < 0.01$). Indeed removal of I1 abolished LTP in our simulations but it could not abolish short-term potentiation (STP) due to transient increase in bouton $Ca^{2+}$ concentration which is present only when astrocyte $Ca^{2+}$ is unblocked (compare Figure 13C and 13D). The transient enhancement in Pr can be observed in figure 13D, but not in figure 13C, which was also observed in the experiments of Perea & Araque (2007). Here we demonstrated that in our model LTP induction is abolished when I1 is removed similar to what has been observed experimentally (Bibb et al, 2001).

Using our computational model we could demonstrate that when ND is coupled with astrocyte stimulation a persistent increase in Pr can be observed. The increase in Pr observed in the experiments of Perea & Araque (2007) was nearly 2.5 times (before ND and astrocyte stimulation: $0.28 \pm 0.04$; after ND and astrocyte stimulation: $0.71 \pm 0.08$) while the increase observed in our model was nearly 1.5 times (before ND and astrocyte stimulation: $0.25 \pm 0.02$; after ND and astrocyte stimulation: $0.35 \pm 0.04$). It should be pointed out that STP observed in the experiments of Perea & Araque (2007), without post-synaptic neuron depolarization, is similar to what we observed in our simulations (experiments: from $0.24 \pm 0.03$ to $0.33 \pm 0.04$; simulations: from $0.26 \pm 0.04$ to $0.33 \pm 0.05$; before and after astrocyte stimulation). We next investigated the possible parameters which could yield us enhancement similar to what was observed in the experiments of Perea & Araque (2007), after post-synaptic neuron depolarization, in neurotransmitter release probability.

Our simulations reveal that persistent increase in neurotransmitter release probability primarily depends upon two parameters which to the best of our knowledge are not yet determined experimentally,

    i)    Percent increase in $Ca^{2+}$ association rate with synaptotagmin ($k_{syt}$), in response to an increase in $PIP_2$ due to NO produced via phosphorylated CaM-KII (discussed in detail in section 2.10), and

ii) The maximal rate at which $IP_3$ is produced by pre-synaptic mGluRs ($v_g$).

We present and discuss results obtained, by varying these two parameters, in the following figures. We first show changes observed in the neurotransmitter release probability (Pr) when we decreased and increased the value of $k_{syt}$. Figures 14A and 14C are for $k_{syt} = 0.025$, and Figures 14B and 14D are for $k_{syt} = 0.1$. Variation in Pr was tested under both the pathways of information processing namely, astrocyte-independent (Figures 14A, B) and astrocyte-dependent (Figures 14C, D).

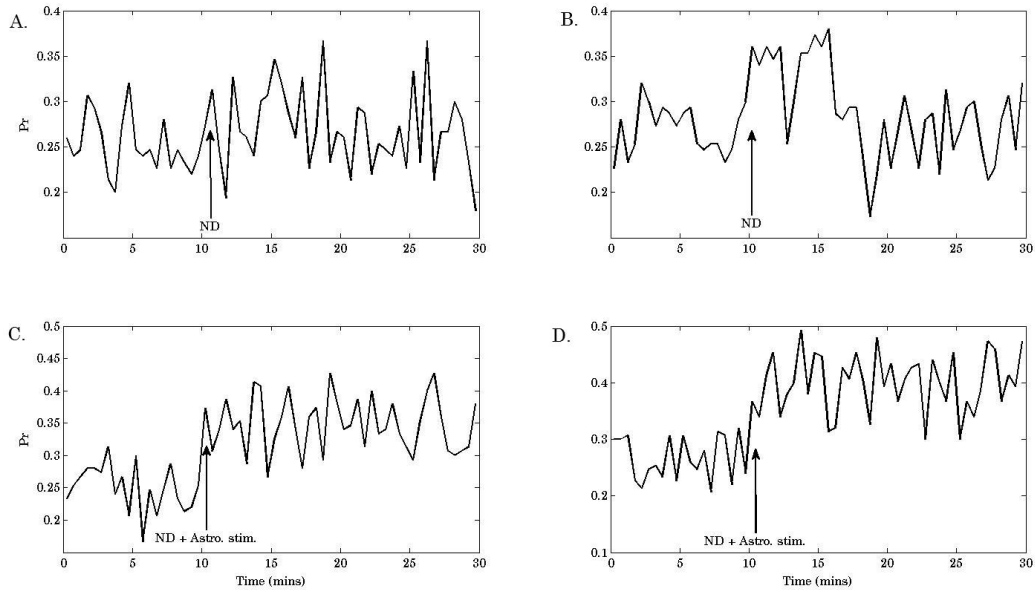

Figure 14. Variation in neurotransmitter release probability (Pr) during astrocyte-independent (A, B) and astrocyte-dependent (C, D) pathway. All the parameters are as listed in Tables 2–9 except $k_{syt} = 0.025$ for A, C and $k_{syt} = 0.1$ for B, D. ND, post-synaptic neuron depolarization. Pr was calculated as previously defined with a time-window of length 30 seconds.

Following suite with our earlier approach, for astrocyte-independent pathway we assumed an NP-EGTA block of astrocyte $Ca^{2+}$ concentration and for astrocyte-dependent pathway we released NP-EGTA block of astrocyte $Ca^{2+}$ concentration at 10 minutes mark (along with post-synaptic neuron depolarization). We tested if the changes in Pr prior to ND and 10 minutes after it are statistically significant using usual two-tailed student's test. As observed earlier in our simulations there was no statistically significant variation in neurotransmitter release probability prior to ND and 10 minutes after it under astrocyte-independent information processing for $k_{syt}$ equal to 0.025 (before: $0.25 \pm 0.03$; 10 minutes after: $0.26 \pm 0.04$; $P = 0.51$; see Figure 14A) and $k_{syt}$ equal to 0.1 (before: $0.27 \pm 0.03$; 10 minutes after: $0.27 \pm 0.03$; $P = 0.83$; see Figure 14B). The results are in agreement with our earlier simulations and experimental findings of Perea & Araque (2007) who found no change in Pr before and after ND when astrocyte was not stimulated. However we found significant change in the neurotransmitter release probability when ND was accompanied with astrocytic $Ca^{2+}$ unblock (see Figures 14C, 14D at the 10 minutes mark) in agreement with experiments of Perea & Araque (2007) and our earlier simulations. The changes observed in Pr before stimulus and 10 minutes after it were statistically significant for $k_{syt}$ equal to 0.025 (before: $0.25 \pm 0.03$; 10 minutes after: $0.35 \pm 0.04$; $P < 0.01$) and $k_{syt}$ equal to 0.1 (before: $0.26 \pm 0.03$; 10 minutes after: $0.4 \pm 0.05$; $P < 0.01$).

Actually varying the percent change in $Ca^{2+}$ association rate with synaptotagmin ($k_{syt}$) mimics the effect of increasing $PIP_2$ concentration over affinity of synaptotagmin for pre-synaptic $Ca^{2+}$ (Bai et al, 2004). For $k_{syt}$ value of 0.025, 0.05 and 0.1 the affinity of synaptotagmin for pre-synaptic $Ca^{2+}$ lies in intervals $[9.75, 10.0]\,\mu M$, $[9.5, 10.0]\,\mu M$ and $[9.0, 10.0]\,\mu M$, respectively. The value of synaptotagmin affinity is the lowest when CaM-KII is fully phosphorylated and the highest when the concentration of the phosphorylated CaM-KII is down and below the threshold for NO production ($P_{1/2}$). Being unaware of the quantitative effects of phosphorylated CaM-KII over synaptotagmin affinity we did not test the change in Pr for values of $k_{syt}$ higher than 0.1.

We next concentrate on the other parameter namely, $IP_3$ production rate by pre-synaptic mGluRs, that enhances neurotransmitter release probability. The maximal rate of $IP_3$ production by pre-synaptic mGluR ($v_g$) can be expressed in terms of surface density of mGluR ($\rho_{mGluR}$), the surface area of bouton exposed to the extra-synaptic glutamate released by astrocyte (S), the Avogadro number $N_A$, the volume of bouton $V_{btn}$, and $IP_3$ molecule production rate per receptor $r_p$ (Nadkarni & Jung, 2007),

$$v_g = \frac{r_p \rho_{mGluR} S}{V_{btn} N_A}. \tag{40}$$

We could not find direct data for surface density of extra-synaptic mGluR on the bouton surface of a CA3 pyramidal cell which is an important parameter determining $v_g$. However, Nusser et al (1995) determined relative densities of synaptic and extra-synaptic $GABA_A$ receptors on cerebellar granular cells using quantitative immunogold method. They estimated that synaptic $GABA_A$ receptors are 230 times more concentrated than that on the extra-synaptic membrane. Holmes et al (1995) suggested synaptic NMDA receptor density to be in the range of 200 – 2000 per $\mu m^2$. Assuming a density of 200 per $\mu m^2$ for synaptic mGluR and using relative densities of synaptic and extra-synaptic ionotropic receptors estimated by Nusser et al (1995). We have an estimate of $\approx 0.87$ per $\mu m^2$ for surface density of extra-synaptic mGluRs on the bouton surface of the CA3 pyramidal cell.

The maximal rate of $IP_3$ production by mGluR on the surface of an astrocyte has been estimated to be 0.062 nM per ms (Nadkarni & Jung, 2007). However, De Pitta et al (2009) used much higher value of (0.2 – 0.5) nM per ms for the maximal rate of $IP_3$ production by mGluR on the astrocyte surface. We assumed the maximal rate of $IP_3$ production by extra-synaptic mGluR to be 0.062 nM per ms i.e. $\frac{0.062 \times 10^{-9} \times 6.023 \times 10^{23}}{10^{-3}}$ molecules/m$^3$/ms or $0.373 \times 10^{17}$ molecules/ms per unit volume. The average volume of a bouton at CA3–CA1 synapse is about 0.13 µm$^3$ (Koester & Sakmann, 2000) which yields the production rate to be ≈0.0048 molecules/ms. If we assume that, IP$_3$ molecule production per receptor is 1 molecule per ms, then the surface exposed to extra-synaptic glutamate ($S$) defines the maximal IP$_3$ production rate. For a maximal rate of IP$_3$ production of 0.062 (nM per ms) $S$ is ≈0.0056 µm$^2$. To study the effect of increasing IP$_3$ production rate ($v_g$) over Pr. We studied change in Pr for an IP$_3$ production rate of 0.1 nM per ms and 0.2 nM per ms. Following a similar process (which is used to determine $S$ for $v_g$ equal to 0.062 nM per ms) it merely implies that for an IP$_3$ production rate of 0.1 nM per ms and 0.2 nM per ms, 0.009 µm$^2$ and 0.018 µm$^2$ of bouton surface should be exposed to extra-synaptic glutamate released by the peri-synaptic astrocyte.

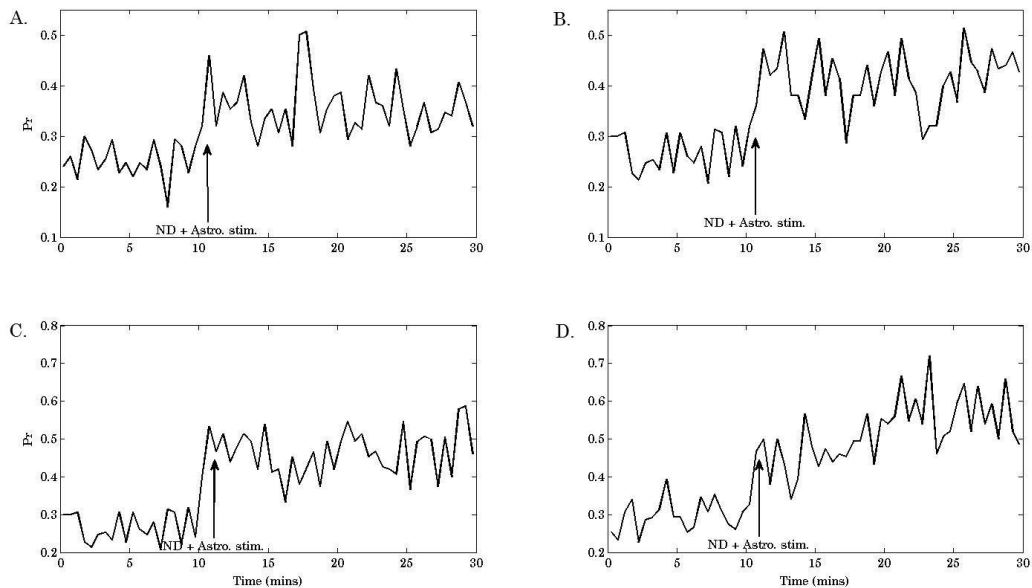

Figure 15. Neurotransmitter release probability (Pr) enhancement using a combination maximal IP$_3$ production rate ($v_g$) and the percent increase in Ca$^{2+}$ association rate to synaptotagmin ($k_{syt}$). A. Is reproduction of figure 9B for comparison purpose, B. For $v_g$ equal to 0.1 nM per ms, C. For $v_g$ equal to 0.1 nM per ms and $k_{syt}$ equal to 0.1, D. For $v_g$ equal to 0.2 nM per ms and $k_{syt}$ equal to 0.1. As may be understood with the text-arrow (↑; ND + Astro. Stim.), all simulations were performed for astrocyte-dependent pathway with post-synaptic neuron depolarization (ND) and astrocyte stimulation occurring at the 10 minutes mark. Note that Y-axis bounds are different for A, B and C, D.

In figure 15, we demonstrate how astrocyte-dependent feed-forward and feed-back pathway is significantly enhanced with variation in combination of $v_g$ and $k_{syt}$. The values of ($v_g$, $k_{syt}$) from Figures 15A–15D are (0.062, 0.05), (0.1, 0.05), (0.1, 0.1) and (0.2, 0.1) respectively. The purpose of the present simulations was to study the quantitative impact of these two parameters over neurotransmitter release probability after the post-synaptic neuron depolarization (ND) and astrocyte stimulation. Indeed we found and can observe a significant increase in Pr from Figures 15A–15D. We calculated Pr, as usual using a time-window of length 30 seconds, and tested Pr before stimulus (ND + astrocyte stimulation) and 10 minutes after it, for statistical significance using two-tailed student's test. Here figure 15A is same as figure 9B but is shown to demonstrate how synaptic coupling is strengthened (see Figures 15B–15D), in our model, by varying only two parameters. The neurotransmitter release probability before stimulus and 10 minutes after it from Figure 15A to Figure 15D increased significantly and is as follows (before: $0.25 \pm 0.03$; after 10 minutes: $0.35 \pm 0.04$; $P < 0.01$), (before: $0.26 \pm 0.04$; after 10 minutes: $0.41 \pm 0.06$; $P < 0.01$), (before: $0.26 \pm 0.04$; after 10 minutes: $0.48 \pm 0.06$; $P < 0.01$) and (before: $0.29 \pm 0.04$; after 10 minutes: $0.57 \pm 0.07$; $P < 0.01$). In Figure 15D we could observe a nearly two-fold increase in pre-stimulus to post-stimulus synaptic activities. The continuous increase in Pr observed from Figure 15D is due to the astrocyte-dependent feed-forward (listening to the synapse) and feed-back loop (talking back to the pre-synaptic neuron). It is interesting to note that almost similar increase in neurotransmitter release probability was also observed in the experiments of Perea & Araque (2007), 10 minutes after pairing of ND and UV-flash astrocyte stimulation (see Figure 4B of Perea & Araque, 2007).

## 4. Conclusion

In this article we investigated LTP of pre-synaptic origin observed in the experiments of Perea & Araque (2007). Here we reveal a possible mechanism by which post-synaptically generated NO can regulate pre-synaptic neurotransmitter release probability. The detail underlying the proposed mechanism is currently experimentally unknown and has been modeled phenomenologically.

Since we were modeling LTP where a possible role of retrograde messenger was suspected, we included a detailed post-synaptic spine model in addition to the detailed pre-synaptic bouton and astrocyte models (see Appendix A for the important progress made in the existing model of Nadkarni et al (2008)). We collected information about the plausible retrograde messenger likely to be responsible for the LTP observed in Perea & Araque (2007). We managed to collect evidences which point to NO as the eluding messenger (we agree that other pathways might well have been involved but NO seems more probable (Alfonso Araque, personal communication)). We eventually found a retrograde signaling pathway backed up with experimental findings see section 2.10 for details, which involves (in sequence of their involvement starting with post-synaptic spine to pre-synaptic bouton) post-synaptic calcium, CaM-KII, NO, pre-synaptic calcium, $PIP_2$ and synaptotagmin. We have also modeled this retrograde signaling process apart from the detailed tripartite synapse.

As we mentioned the questions to be answered were more associated with post-synaptic spine. Not much literature is available as to how exactly the pre-synaptic calcium stores are regulated by astrocytic glutamate. Therefore the classic pathways involving agonist (astrocytic glutamate) dependent $IP_3$ production and $IP_3R$ mediated calcium release have been modeled. The vesicle based model proposed for glutamate release from astrocytes incorporates latest advances in glial pathophysiology (Malarkey & Parpura, 2011) which reveal the number of releasable synaptic-like microvesicles (SLMVs) inside astrocytes and also the percentage (or probability) of SLMVs released in response to a mechanical stimulation. Both these parameters, especially the number of releasable SLMVs decides the effect of astrocyte over pre-synaptic calcium concentration and consequently on synaptic plasticity.

We tried to quantify the enhanced pre-synaptic neurotransmitter release probability observed in the experiments by Perea & Araque (2007) after the post-synaptic neuron depolarization and peri-synaptic astrocyte stimulation. Our computational study predicts two important parameters ($k_{syt}$ and $v_g$) which can enhance neurotransmitter release probability and consequently synaptic efficacy at CA3–CA1 synapses. After slight modification in both the parametric values we could observe an increase in Pr (neurotransmitter release probability from pre-synaptic bouton) similar in amplitude to their experiments (simulations: from $0.29 \pm 0.04$ to $0.57 \pm 0.07$; experiments: from $0.28 \pm 0.04$ to $0.71 \pm 0.08$; before and after astrocyte stimulation). However one thing should be mentioned here that in their experiments Perea & Araque (2007) used UV-flash photolysis to stimulate astrocytes independently of the pre-synaptic neuron activity. Further the use of UV-flash could have stimulated astrocyte $Ca^{2+}$ to an extent which might not be possible under physiological conditions (i.e. in an activity dependent manner) and hence observed a slightly higher neurotransmitter release probability in their experiments. Both these parameters are experimentally unknown at CA3–CA1 synapses. However using our modeling approach we gave an estimate for these parameters, which is in conformity with the data from the other synapses (Nusser et al, 1995; Holmes, 1995; Bollman et al, 2000).

Through computer simulations of the proposed model we demonstrate that our assumption, NO-production threshold is crossed only under astrocyte-dependent pathway, holds for under different parametric values (see Figure 11). The validity and range of such a NO-production threshold is open for experimental testing. We also found that astrocytes help in potentiating hippocampal synapses in a feed-forward and feed-back manner, where astrocytes sense the signal transduced by the pre-synaptic neuron and feeds it back into the pre-synaptic neuron resulting in strengthening of the synapse (see Figure 15D where such a phenomenon is clearly visible and is in close agreement with the experiments of Perea & Araque (see Figure 4B of Perea & Araque, 2007)). Following Perea & Araque (2007) we used single pulses to stimulate pre-synaptic terminal and found that the astrocytes potentiate CA3–CA1 tripartite synapses. However we also used 0.33 Hz TBS (Epochs separated at an interval of 3 seconds with 1 epoch equal to 4 pulses at 100 Hz) and found an increase in synaptic efficacy (see Figure S4, provided with the electronic supplementary material) when the innervating astrocyte was turned on (but we first adjusted $\rho_{Ca}$ so that Pr remains between 0.2–0.3 in absence of an astrocyte).

The proposed model is versatile and can exhibit STP as well, if we do not depolarize the post-synaptic neuron, similar to the experiments of Perea & Araque (2007). The model was shown to exhibit STP, in a special case, where I1 was assumed to be removed from the PSD (see Figure 13). One more important aspect of our model is that it reaches steady-state. This mean that it continues to operate in a realistic manner with repetitive stimulations and comes to its resting conditions when not stimulated. Therefore the model is appropriate for use in situations where continued use for longer periods is necessary (e.g. LTP).

To conclude, we have formulated a realistic mathematical model of LTP (of pre-synaptic origin) at a SC–CA1 synapse modulated by an astrocyte. This model uses recent advances in astrocyte physiology which aims to classify vesicular release of glutamate from the astrocytes (Malarkey & Parpura, 2011). It helped us to quantify glutamate release from the astrocyte and to improve over previously proposed models (Volman et al, 2007; Nadkarni et al, 2007; Nadkarni et al, 2008). This model should be useful in simulating signal transduction under variety of pharmacological and pathophysiological conditions which are demonstrable in the laboratory. It will allow in-depth analysis of mechanisms behind these processes, particularly when experimental data are hard to obtain.

**Acknowledgments**

The work has been supported by the Department of Science and Technology, Government of India, grant no. SR/CSI/08/2009. Helpful discussions and suggestions from Prof. Alfonso Araque, Instituto Cajal, Spain is being thankfully acknowledged.

**Appendix A**

The idea of the astrocytic feedback into the pre-synaptic terminal is not new as it has been previously modeled by Nadkarni et al (2008). However our model is new and more challenging in the sense that it is biologically more detailed than Nadkarni et al (2008). For example, Nadkarni et al (2008) modeled a CA3–CA1 pyramidal cell (CA3–CA1) synapse as modulated by an astrocyte. It needs to be pointed out that the parameter values chosen by them for bouton calcium in response to an action potential (AP) are biologically not valid at the CA3–CA1 synapses. They assume that the intracellular calcium concentration rises instantly to 300 μM (and falls back to the resting values after 2 ms) in response to an AP. If we assume such high concentration then it implies that nearly $\frac{0.13 \times 300 \times 10^{-6} \times 6.023 \times 10^{23}}{10^{15}} \approx 23490$ ions (for an average pyramidal cell bouton of volume 0.13 μm$^3$, Koester & Sakmann, 2000) are free per AP. However the total calcium charge entering into a **hippocampal pyramidal cell bouton** during an AP is less than 1 fC (femto Coulomb) (Koester & Sakmann, 2000). It simply implies that the total calcium ions entering a hippocampal pyramidal bouton during an AP is less than $\frac{1 \times 10^{-15}}{2 \times 1.6 \times 10^{-19}} = 3125$ ions which is considerably much less than 23490 ions as discussed before. In our case, calcium entering through an AP is around 5 μM i.e.

$$\frac{0.13 \times 5 \times 10^{-6} \times 6.023 \times 10^{23}}{10^{15}} \approx 392$$ ions which is well under 3125 ions (maximum number of calcium ions entering into a hippocampal pyramidal bouton).

Furthermore, Weber et al (2010) concluded that the neurons with extracellular calcium concentration of 2 mM are unlikely to have calcium sensor affinity as high as 100 μM which makes the choice of Bertram model unfavorable (since its four sensors or sites have affinities 108 nM, 400 nM, 200 μM and 1334 μM) at least for the modeling of the pre-synaptic bouton neurotransmitter release. Instead of high calcium sensor affinity Weber et al (2010) suggested 10 μM as the best estimate for calcium sensor affinity which has been used in our model.

The astrocytic calcium dynamics used in Nadkarni et al (2008) and in our model may seem identical in broad sense (since both of them have an agonist-dependent inositol triphosphate ($IP_3$) production term, agonist-independent $IP_3$ production term and $IP_3$ degradation term), but they are not. For example, the calcium-dependent term used in their model is based on the classic De Young & Keizer (1992) model which talks about phospholipase C (PLC) (which is activated by calcium) dependent $IP_3$ production is based on data from the experiments over the WRK1 cells (Mouillac et al 1990), liver cells (Taylor & Exton, 1987) etc. On the other hand we make use of the G-ChI model (De Pitta et al, 2009), which has a more detailed $IP_3$ degradation term (it incorporates inositol polyphosphate 5-phosphatase (IP-5P) and $IP_3$ 3-kinase ($IP_3$-3K) based $IP_3$ degradation), agonist-dependent term (it incorporates $Ca^{2+}$/protein kinase C (PKC)-dependent inhibitory factor over agonist-dependent $IP_3$ production term) and agonist-independent $IP_3$ production term (it incorporates PLCδ-dependent $IP_3$ production term) based on an astrocyte specific experiment.